\documentclass{pasa}%

\usepackage{graphicx}

\title[MWA Beam Model]{Calibration and Stokes Imaging with Full Embedded Element Primary Beam Model for the Murchison Widefield Array}
\author[Marcin Sokolowski et al.]
{M.~Sokolowski$^{1,2,}$\thanks{marcin.sokolowski@curtin.edu.au}, T.~Colegate$^1$, A.~T.~Sutinjo$^1$, D.~Ung$^1$, R.~Wayth$^{1,2}$, N.~Hurley-Walker$^1$, E.~Lenc$^{2,3}$, B.~Pindor$^{2,4}$,
J.~Morgan$^1$, D.~L.~Kaplan$^{5}$, M.~E.~Bell$^{2,6}$, J.~R.~Callingham$^{2,3,6}$, K.~S.~Dwarakanath$^7$, Bi-Qing~For$^8$,
B.~M.~Gaensler$^{2,3,9}$, P.~J.~Hancock$^{1,2}$, L.~Hindson$^{10}$, M.~Johnston-Hollitt$^{10,11}$, A.~D.~Kapi\'nska$^{2,8}$, 
B.~McKinley$^{2,4}$, A.~R.~Offringa$^{12}$, P.~Procopio$^{2,4}$, L.~Staveley-Smith$^{2,8}$, C.~Wu$^{8}$, Q.~Zheng$^{10,11}$\\
\affil{$^1$International Centre for Radio Astronomy Research, Curtin University, GPO Box U1987, Perth, WA 6845, Australia}%
\affil{$^2$ARC Centre of Excellence for All-sky Astrophysics (CAASTRO), Redfern, NSW, Australia}%
\affil{$^3$Sydney Institute for Astronomy, School of Physics, The University of Sydney, NSW 2006, Australia}
\affil{$^4$School of Physics, The University of Melbourne, Parkville, VIC 3010, Australia}
\affil{$^5$Department of Physics, University of Wisconsin--Milwaukee, Milwaukee, WI 53201, USA}
\affil{$^6$CSIRO Astronomy and Space Science, Marsfield, NSW 2122, Australia}
\affil{$^7$Raman Research Institute, Bangalore 560080, India}
\affil{$^8$International Centre for Radio Astronomy Research, University of Western Australia, Crawley 6009, Australia}
\affil{$^9$Dunlap Institute for Astronomy and Astrophysics, 50 St. George St, University of Toronto, ON M5S 3H4, Canada}
\affil{$^{10}$School of Chemical \& Physical Sciences, Victoria University of Wellington, Wellington 6140, New Zealand}
\affil{$^{11}$Peripety Scientific Ltd., PO Box 11355 Manners Street, Wellington 6142, New Zealand}
\affil{$^{12}$Netherlands Institute for Radio Astronomy (ASTRON), PO Box 2, 7990 AA Dwingeloo, The Netherlands}
}
\jid{PASA}
\doi{10.1017/pas.\the\year.xxx}
\jyear{\the\year}

\usepackage{aas_macros}
\usepackage{hyperref} 
\hypersetup{colorlinks,citecolor=blue,linkcolor=blue,urlcolor=blue}

\newcommand{\red}[1] {#1}
\newcommand{\rednew}[1] {#1}

\usepackage{amsmath}
\usepackage{gensymb}
\usepackage{float}
\usepackage[caption = false]{subfig}

\begin{document}%

\begin{frontmatter}
\maketitle
\begin{abstract}
The Murchison Widefield Array (MWA), located in Western Australia, is one of the low-frequency precursors of the international Square Kilometre Array (SKA) project.
In addition to pursuing its own ambitious science program, it is also a testbed for wide range of future SKA activities ranging from hardware, software to data analysis.
The key science programs for the MWA and SKA require very high dynamic ranges, which challenges calibration and imaging systems.
Correct calibration of the instrument and accurate measurements of source flux densities and polarisations require precise characterisation of the telescope's primary beam.
Recent results from the MWA GaLactic Extragalactic All-sky MWA (GLEAM) survey show that the previously implemented Average Embedded Element (AEE) model still leaves residual polarisations errors of up to 10-20\% in Stokes Q.
\red{We present a new simulation-based Full Embedded Element (FEE) model which is the most rigorous realisation yet of the MWA's primary beam model. It enables efficient calculation of the MWA beam response in arbitrary directions without necessity of spatial interpolation.}
In the new model, every dipole in the MWA tile ($4\times4$ bow-tie dipoles) is simulated separately, taking into account all mutual coupling, ground screen and soil effects, and therefore accounts for the different properties of the individual dipoles within a tile.
\red{We have applied the FEE beam model to GLEAM observations at 200--231\,MHz and used false Stokes parameter leakage as a metric to compare the models. We have determined that the FEE model reduced the magnitude and declination-dependent behaviour of false polarisation in Stokes Q and V while retaining low levels of false polarisation in Stokes U.}
\end{abstract}

\begin{keywords}
instrumentation: interferometers 
\end{keywords}
\end{frontmatter}
\section{INTRODUCTION }
\label{sec:intro}

Accurate astronomical measurements with radio interferometric telescopes require correction of instrumental effects. For fixed-antenna aperture array telescopes, 
the primary beam of the interferometer elements varies considerably with pointing direction. It is important to accurately model and correct for the primary beam, 
as differences between the actual and modelled beam result in errors during both telescope calibration and imaging.
\red{Moreover, correct calibration of the primary beam effects (i.e. beam chromaticity and polarisation leakage) in low-frequency aperture arrays is critical for detection of the Epoch of Reionisation (EoR), which is a key science goal of the Square Kilometre Array (SKA)\footnote{http://www.skatelescope.org/} and its precursors. \citet{asad_et_al_2016_lofar} studied beam effects and their impact on the EoR science for the Low Frequency Array (LOFAR)  and analysis for the Hydrogen Epoch of Reionization Array (HERA) can be found in \citep{2017PASP..129d5001D}. Here we present the new primary beam model for the Murchison Widefield Array (MWA), which is a low-frequency ($\sim$75-300\,MHz) telescope and SKA precursor that commenced scientific operation in 2013 \citep{TinGoe13}. The challenges of low-frequency radio polarimetry, based on experiences from the MWA, were summarised in the recent paper by \citet{2017arXiv170805799L}.}

\begin{figure}
\begin{center}
\includegraphics[width=\columnwidth]{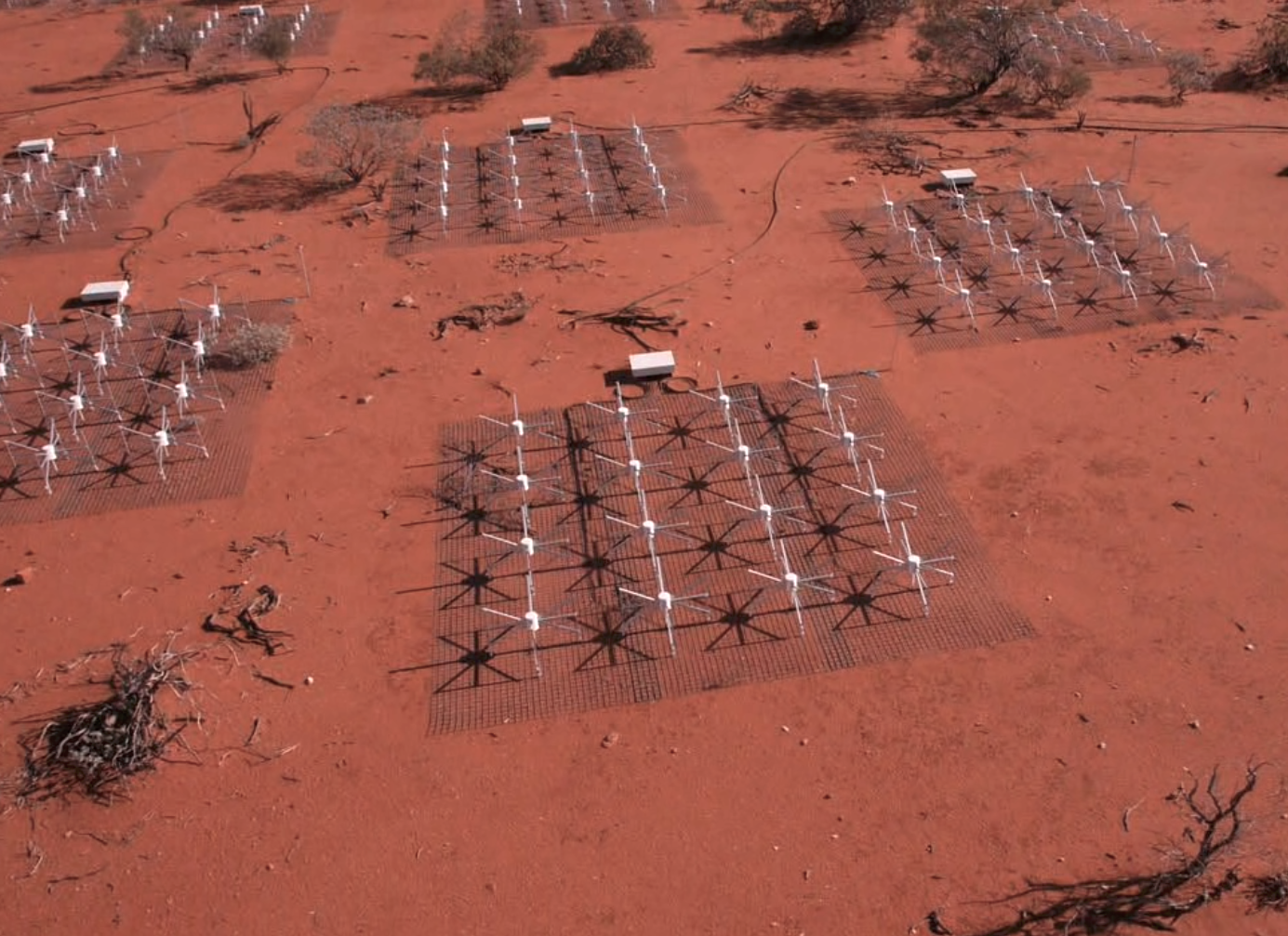}
\caption{Examples of the MWA's aperture array antenna `tiles', each comprised of a $4 \times 4$ grid of individual bow-tie dipoles (Image credit: MWA Project, Curtin University).}
\label{fig:MWA-tile}
\end{center}
\end{figure}

The Phase I deployment of the MWA consists of 128 `tiles' with separations up to 3\,km, each tile being a small electronically steerable phased array of 16 dual-polarised bow-tie antennas (Fig. \ref{fig:MWA-tile}).
The steering is provided by beamformer units where appropriate 5 bit time delays (in discrete steps of 435\,ps) are applied to each of the 16 antennas in the tile. 
The signals from each tile are digitised and cross-correlated with other tiles, and the visibility measurements archived.
The archived data covers 24 coarse channels of 1.28\,MHz, which are divided into fine channels of 10 or 40\,kHz - depending on the correlator settings \citep{steve_ord}.

With processing pipelines now in place for key science observing with
the MWA, such as detecting emission from the epoch of re-ionisation
(EoR) \citep{2016ApJ...825..114J} and the GaLactic Extragalactic All-sky MWA (GLEAM) survey
\citep{GLEAM_Wayth,gleam_nhw}, focus has turned to the accuracy of the primary beam
models for the MWA.

An incorrect beam model manifests itself during calibration and imaging
of the target field. A simple observing scenario is where the tile-based
complex gain calibration solutions are determined by observing a strong,
point-like source.
Ideally, these calibration solutions are corrected for direction-dependent
effects (DDEs, i.e., the modelled gain of the tile beam in the direction
of the calibrator source), resulting in measurement of the direction-independent electronic gain for each tile.
These calibration solutions can then be applied as a correction to the target visibility data collected on a field with or without a suitable calibrator source.
In addition, the DDEs in the target pointing direction
must again be corrected for.
Errors in the beam model propagate through as errors in the DDE-corrected observation. 
The level of error will depend on the pointing direction of the tile beam during calibration and observation, and the location of the sources within the beam.
The most sensitive probe of model inaccuracies are polarimetric measurements of the celestial sources because errors in the beam model for linearly polarised receptors, like the MWA's tiles, manifest themselves as false Stokes in the calibrated images \citep{lenc_et_al_2016}.

In response to false Stokes Q observed in the data calibrated with a simple analytical (\rednew{Hertzian} dipole) tile beam model, a new, FEKO \footnote{from German \textbf{FE}ldberechnung bei \textbf{K}\"orpern mit beliebiger \textbf{O}berfl\"ache which can be translated as ``field calculations involving bodies of arbitrary shape''.} software simulation based, model was implemented using an ``average embedded element'' (AEE) pattern \citep{SutOSu15}. 
The AEE model showed significant improvements with respect to analytical model and reduced false Stokes Q in the calibrated data from $\approx$30\% to typically below 10\%.
The AEE model was used to calibrate the GLEAM survey. However, a noticeable (5-20\%) false Stokes Q is still reported in the GLEAM calibrated data, which is attributed to the AEE beam model \red{(see Fig.~4 in \citet{gleam_nhw} and Fig.~\ref{fig:ratio_Sxx_div_Syy} in this paper)}. 
Furthermore, \citet{offringa-2016} attribute $\sim$1\% leakage into Stokes V to inaccuracies in the beam model, but the false Stokes V leakage can be higher on individual snapshot images (not averaged over long periods of time).
Therefore, further improvements in the MWA beam model are required to improve the accuracy of the calibration and polarimetric measurements.

This paper is organised as follows. In Section~\ref{sec_model_desc} we present implementation of the new Full Embedded Element (FEE) primary beam model of MWA. We describe the physical representation of the tiles, the mathematical description in terms of spherical harmonics and summarise differences with respect to the previous AEE model. 
In Section~\ref{sec_beam_corr_proc} we present a beam correction procedure that we implemented to test the new beam model.
In Section~\ref{sec_results} we present the results of this procedure applied to MWA data and compare performances of the new FEE model against the previous AEE model.

\section{MODEL DESCRIPTION}
\label{sec_model_desc}

\citet{SutOSu15} proposed three tiers of beam modelling sophistication: 1) an analytic model using array theory and pattern multiplication, 2) using the AEE and incorporating mutual coupling, as detailed in that paper, and 3) using an FEE model.
The improvements from the AEE model were incorporated into the mainstream MWA data processing pipelines.
Here we improve the model again by using the FEE patterns for each of the bow-tie antennas to model the tile beam.

From the MWA user perspective, the approach is consistent with previous
models where the beam pattern is generated on-the-fly for a given
set of antenna delays\footnote{The MWA's analogue beamformers use true time delays to generate frequency independent beams rather than phase shifts, which are only valid over a narrow fractional bandwidth} which define the pointing direction of the beam.
Due to computational limitations the embedded element pattern is calculated for the centre of every 1.28\,MHz coarse channel. At arbitrary frequency channels the beam model is calculated for the closest coarse channel which is within 0.64\,MHz (i.e. no frequency interpolation was implemented at the current version of the beam model, but it can be added in future if even higher precision is required).
The internals are different to previously implemented models in three key areas:
\begin{itemize}
\item The physical model of the tile is improved (i.e. size of the ground screen now reflects the actual 5x5\,m mesh size).
\item For the first time, the beam pattern is calculated using the full
embedded element patterns of the 16 bow-tie dipoles in the tile-dipole.
\item The beam pattern is calculated using spherical wave expansion, allowing
accurate replication of full wave simulation of the beam pattern in any direction.
\end{itemize}
We describe these differences in more detail below.

\subsection{Physical model of the tile}

An MWA tile comprises of 16 bow-tie antennas aligned in the east-west
($x$) and north-south ($y$) directions, located on a regular $4\times4$
grid with 1.1\,m spacing between centres, as shown in Fig.~\ref{fig:MWA-tile}.
The normal operating frequency range is 75--240\,MHz, but is capable
of observations up to $\sim$315~MHz. \citet{TinGoe13} and \citet{NebHew16}
give a detailed description of the physical tile. 

\begin{figure}[!t]
\centering{}\includegraphics[bb=0bp 0bp 1000bp 829bp,width=9cm]{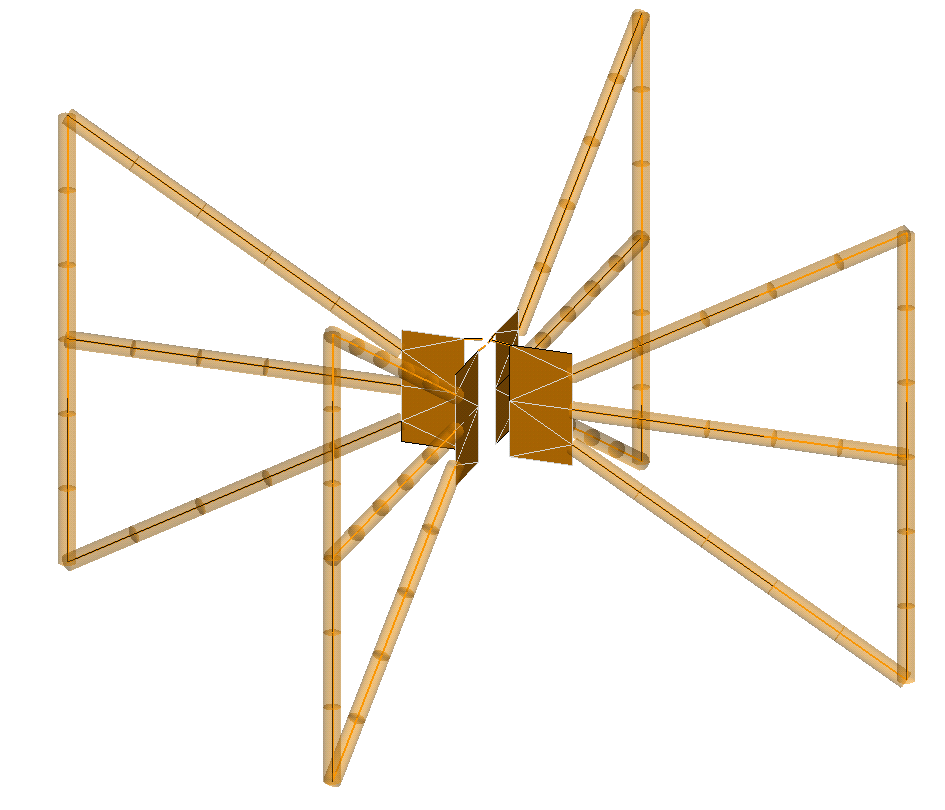}
\caption{The bow-tie antenna modelled in FEKO. The marks along the arms indicate the simulation segments.}
\label{fig:FEKO-bow-tie}
\end{figure}

Figure~\ref{fig:FEKO-bow-tie} shows the bow-tie antenna modelled in
the FEKO\footnote{\url{www.feko.info}} electromagnetic simulation software. It
has the same dimensions as the actual antenna, but uses a\textbf{
}15~mm\textbf{ }diameter wire instead of the aluminium channels to
reduce simulation complexity. This diameter was selected because it
best matches the measured \red{dimensions of the antenna elements}. The ground
mesh has spacing of 5~cm between adjacent wires, which at MWA frequencies
we model as a perfect electrical conductor (PEC). Beyond the extent
of the $5\times5$~m ground plane we simulate the ground as soil from the Murchison Radio-astronomy Observatory (MRO)
with 2\% moisture, based on the permittivity and conductivity properties
of soil from the MRO reported in \citet{SutCol15}. Antennas are elevated
by 10\,cm from the mesh, but the plastic legs were not included in the simulation. 

\begin{figure}[!t]
\centering{}\includegraphics[width=8cm]{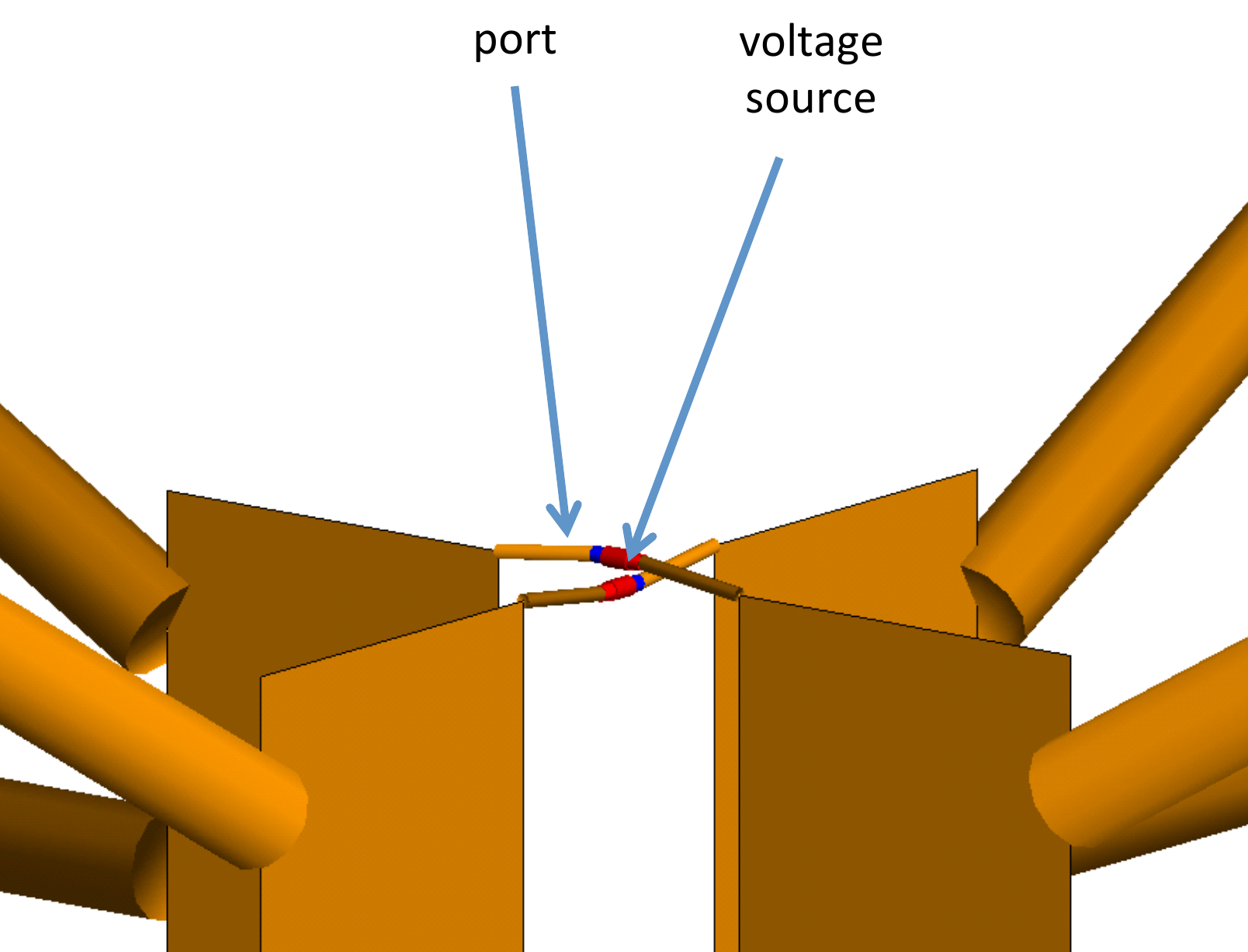}

\caption{Bow-tie ports modelled in FEKO, where the ports are loaded with a voltage source.}
\label{fig:FEKO-bow-tie-zoom}
\end{figure}

\begin{figure}[!t]
\centering{}
\includegraphics[width=\columnwidth]{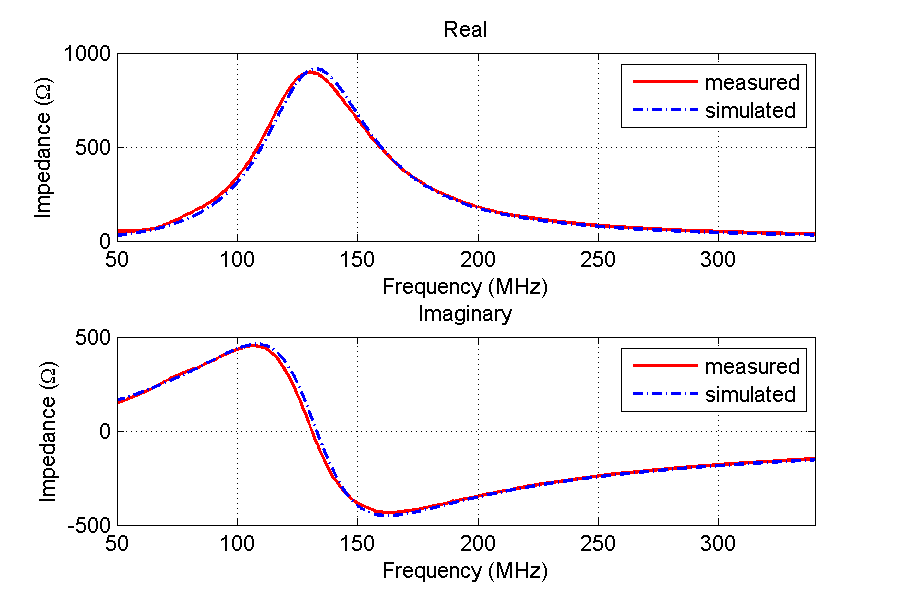}
\caption{Comparison between the measured LNA impedance and lumped circuit model (simulated in FEKO). We have verified that the small residual difference between the measured and model impedance has an insignificant \rednew{(< 1\%)} impact on the resulting beam model \rednew{at 216\,MHz}.}
\label{fig:lumped-circuit}
\end{figure}

The model uses loaded ports as shown in Fig.~\ref{fig:FEKO-bow-tie-zoom}.
The low noise amplifier (LNA) impedance is modelled using a lumped circuit, meaning the complex
characteristics of the LNA impedance is predictable by measuring a
simple shunt-series elements made up of resistors, inductors and capacitors
(RLC). We represent the LNA input impedance with a 2\,nH series element
attached to a RLC (914 $\Omega$, 450 nH and 3.2 pF) shunt network.
As per Fig.~\ref{fig:lumped-circuit}, this model shows good agreement
with the LNA impedance measured\textbf{ }using a vector network analyser.
The lumped circuit model results in a more compact, self-contained
simulation file, and returns an impedance at arbitrary frequency resolution.

\subsection{Jones matrix beam model}

The primary beam of aperture array telescopes strongly varies with
pointing direction, and differs between the orthogonal antenna polarisations
\emph{x} and \emph{y}. At a particular pointing direction $\theta$
(angle from zenith) and $\phi$ (azimuth angle, increasing clockwise
from north through east), we can describe the instrumental effect
of the MWA tile on the astronomical signal as a Jones matrix $\mathbf{J}$,
a $2\times2$ complex matrix. This Jones matrix maps the voltage at
the beamformer output (\emph{$v_{x}$} and \emph{$v_{y}$}) to the
signal in orthogonal sky polarisations ($e_{\theta}$ and $e_{\phi}$)
as follows: $\mathbf{v}=\mathbf{J}\mathbf{e}$, which in expanded
form is
\begin{equation}
\begin{bmatrix}\begin{array}{c}
v_{x}\\
v_{y}
\end{array}\end{bmatrix}=\begin{bmatrix}\begin{array}{cc}
J_{x\theta} & J_{x\phi}\\
J_{y\theta} & J_{y\phi}
\end{array}\end{bmatrix}\begin{bmatrix}\begin{array}{c}
e_{\theta}\\
e_{\phi}
\end{array}\end{bmatrix},
\end{equation}
where $v_{x}$ and $v_{y}$ are the voltages from the $x$ (E-W) and
$y$ (N-S) measurement bases (bow-tie antennas), and $e_{\theta}$
and $e_{\phi}$ are far-field unit vectors in spherical coordinate
bases \citep{Smi11-I,SutOSu15}.

We can separate the MWA tile Jones matrix into two principal components:
\begin{equation}
\mathbf{J}=\mathbf{GE},
\label{eq:J-separate}
\end{equation}
where $\mathbf{G}$ is the direction-independent effect (DIE) due
to complex electronic gain and $\mathbf{E}$ is the direction-dependent
effect (DDE) due to the tile beam pattern. The latter varies as a
function of the tile pointing (amplitude and phase weights of each
antenna), and, for a given pointing, a Jones matrix applies for each
($\theta,\phi$) point in the hemisphere. In FEKO, $\mathbf{E}(\theta,\phi)$
can be calculated for a given pointing by applying an appropriate
phase slope across the tile. However, such an approach would require
a new simulation run for every desired pointing. Instead we model $\mathbf{E}(\theta,\phi)$ for each embedded element, being the in-situ
radiation pattern of each bow-tie dipole in the tile \citep{fee_kelley_stutzman}. 
Each antenna is simulated in turn by setting its amplitude to a constant voltage,
and the other 15 to zero. The resulting Jones matrix is normalised to the zenith ($\theta=0)$ of a zenith-pointed beam (zero delay, 
thus no phase slope across the tile), therefore the absolute value of the excitation voltage is not important, as long as it is equal between antennas. 
The tile beam pattern for a given pointing can be determined post-simulation
by weighting each element pattern with the appropriate phase and amplitude.

\subsection{Computational representation via Spherical Wave Expansion }

Different methods are possible to represent the full wave simulation
results from FEKO in the beam model. The method used previously for the AEE model was to output $\mathbf{E}$ for regular $(\theta,\phi)$
intervals on the hemisphere. The disadvantage of this approach is
that interpolation is required between adjacent $(\theta,\phi)$ points.
In this paper, we use Spherical Wave Expansion (SWE) to calculate the tile beam pattern
from the electric far-field according to the following formula \citep[chapter ``AS card'']{feko_manual}:

\begin{equation}
\begin{split}
\overrightarrow{E}^{\mathrm{ff}}(\theta,\phi) =
\beta\sqrt{\frac{Z_{0}}{2\pi}}\frac{e^{-j\beta }}{\beta} \times \hspace{2cm} \\
 \Bigg[\sum_{n=1}^{\infty}\sum_{m=-n}^{n}\frac{e^{jm\phi}C_{mn}}{\sqrt{n(n+1)}}
 \Big(\frac{-m}{\left|m\right|}\Big)^{m} 
 \big( e_{mn}^{\theta}\hat{\theta}+e_{mn}^{\phi}\hat{\phi}\big)
 \Bigg],
\label{eqn:E_ff}
\end{split}
\end{equation}

where $\beta$ is the wavenumber, $Z_{0}$ is the intrinsic impedance of free space, $C_{mn}= ((2n+1)(n-|m|)!)/(2(n+|m|)!)^{1/2}$
is the normalisation factor for the associated Legendre function, $P_{n}^{|m|}(\cos\theta)$, of order $n$ and rank $|m|$ \citep[see Chapter 6 in ][]{Harrington_2001_ch6}.
The coefficients $e_{mn}^{\theta}$ and $e_{mn}^{\phi}$ can be calculated according to the following equations:
\begin{equation}
\raisetag{6pt}
\begin{split}
         e_{mn}^{\theta}=\Bigg[j^{n}\frac{P_{n}^{|m|}(\cos\theta)}{\sin\theta}\Big(|m|Q_{2mn}^{tile}\cos\theta-mQ_{1mn}^{tile}\Big) \\
         + j^{n}Q_{2mn}^{tile}P_{n}^{|m|+1}(\cos\theta)\Bigg],
\raisetag{20pt}
\label{eqn:small_e_thB}
\end{split}
\end{equation}

\begin{equation}
\begin{split}
         e_{mn}^{\phi}=\Bigg[j^{n+1}\frac{P_{n}^{|m|}(\cos\theta)}{\sin\theta}\Big(mQ_{2mn}^{tile}-|m|Q_{1mn}^{tile}\cos\theta\Big) \\
         -j^{n+1}Q_{1mn}^{tile}P_{n}^{|m|+1}(\cos\theta)\Bigg], 
\label{eqn:small_e_phiB}
\end{split}
\end{equation}
where $s=1$ and $s=2$ in $\mathbf{Q}_{smn}^{tile}$ refer to transverse electric (TE) and transverse magnetic (TM) modes respectively. $\mathbf{Q}_{smn}^{tile}$ are vectors of coefficients formed as linear combination of 16 embedded elements beam patterns for 
specific pointing $(\theta,\phi)$ according to beamformer time delays $t_i$ and $\mathbf{Q}_{smn}^i$ are FEKO-generated coefficients for every $i$-th antenna in the MWA tile. Hence, vectors $\mathbf{Q}_{smn}^{tile}$ can be calculated as:
\begin{equation}
   \mathbf{Q}_{smn}^{tile} = \sum_{i=1}^{16} e^{-2 \pi j\nu t_{i}} \mathbf{Q}_{smn}^{i},
\label{eqn:q_tile}
\end{equation}
where $\nu$ is the observing frequency. 
The spherical harmonics approach allows rapid and accurate computation
of the beam at one or more desired ($\theta,\phi$) coordinates. 

\red{We have implemented the new FEE beam model in Python for MWAtools (an internal software package for MWA data processing) and in C/C++. The C/C++ implementation is included in the Real-Time System (\textsc{rts}) \citep{RTS_4703504} and its off-line implementation by \citep{offringa-2016} known as \textsc{calibrate}.
The C/C++ implementation can calculate the beam model for 1 million spatial points ($1000\times1000$ pixels) in a few minutes. However, the Python implementation requires spatial interpolation in the beam calculation to generate similarly sized beam models within a similar time-frame.
}

\section{FULL-STOKES BEAM CORRECTION PROCEDURE}
\label{sec_beam_corr_proc}

\emph{}The visibility matrix for the cross-correlation of tiles $i$
and $j$ is measured as
\begin{equation}
\mathbf{V}_{ij}=2\begin{bmatrix}\begin{array}{cc}
\langle v_{i,x}v_{j,x}^{*}\rangle & \langle v_{i,x}v_{j,y}^{*}\rangle\\
\langle v_{i,y}v_{j,x}^{*}\rangle & \langle v_{j,x}v_{j,x}^{*}\rangle
\end{array}\end{bmatrix}=2\begin{bmatrix}\begin{array}{cc}
{\rm XX} & {\rm XY}\\
{\rm YX} & {\rm YY}
\end{array}\end{bmatrix},\label{eq:Bapp-cross-corr}
\end{equation}
where $v$ is the voltage from the beamformer of a given tile (i or j) and antenna polarisation (x or y). 
In the calibration observation scenario, a single bright source dominating the visibilities is observed
and hence the brightness matrix $\mathbf{B}$ represents this single point source (delta function).
Therefore in the calibration scenario, for an ideal cross-correlator, $\mathbf{V}_{ij}$ is related to the brightness matrix $\mathbf{B}$ of the actual ``single strong source sky'' by the response of each tile, represented as a Jones matrix $\mathbf{J}$ (eq.~\ref{eq:J-separate})\footnote{The general relation involving integral over the sky can be found in radio astronomy textbooks \cite{THOMPSON} or \cite{1999ASPC..180.....T}}: 
\begin{equation}
\mathbf{V}_{ij}=\mathbf{J}_{i}\mathbf{B}\mathbf{J}_{j}^{\mathbf{H}},\label{eq:Vij}
\end{equation}
where $\mathbf{B=}\langle\mathbf{ee}^{{\rm H}}\rangle$ and the \textbf{$\mathbf{H}$}
superscript denotes the Hermitian transpose. Using equation~\ref{eq:J-separate}, we can separate $\mathbf{J}$ as two Jones matrices:
\begin{equation}
\mathbf{V}_{ij}=\mathbf{G}_{i}\mathbf{E}_{i}\mathbf{B}\mathbf{E}_{j}^{\mathbf{H}}\mathbf{G}_{j}^{\mathbf{H}},\label{eq:Vij-expanded}
\end{equation}
where \textbf{$\mathbf{G}$ }describes the direction-independent effects
due to complex electronic gain and $\mathbf{E}$ the direction-dependent
effects due to the tile beam pattern. This layered description of
effects on the signal path is known as the ``radio interferometer
measurement equation'' \citep{HamBre96-I,Smi11-I}.

If the direction-independent effects ($\mathbf{G}$ in eq.~\ref{eq:J-separate})
are correctly calibrated for, we observe what \citet{Smi11-II} calls
the ``apparent sky'', being the true sky attenuated by the tile
beam \textbf{$\mathbf{E}$}: 
\begin{equation}
\mathbf{B_{{\rm app}}}=\mathbf{E}_{i}\mathbf{B}\mathbf{E}_{j}^{\mathbf{H}}.\label{eq:Bapp}
\end{equation}
If we also assume identical beam patterns for tiles $i$ and $j$,
the sky brightness matrix can be estimated using a model Jones matrix
$\mathbf{\widetilde{E}}$ representing the MWA tile beam:
\begin{equation}
\widetilde{\mathbf{B}}=\mathbf{\widetilde{E}^{-1}B_{{\rm app}}}\left(\mathbf{\widetilde{E}}^{H}\right)^{-1},\label{eq:B-est}
\end{equation}
where the tilde designates a modelled matrix or a result estimated from models. \red{We will use this convention through the remainder of the paper.}

We note that the assumption of identical beam patterns for all tiles significantly simplifies the data processing, since corrections to images can all be applied in image-space, as a linear combination of images made in instrumental polarisation coordinates.

From $\widetilde{\mathbf{B}}$, we can calculate Stokes parameters (following the convention of Smi11-I)\emph{:}
\begin{align}
\widetilde{I} & =(\widetilde{B}_{1,1}+\widetilde{B}_{2,2})/2\label{eq:Stokes}\\
\widetilde{Q} & =(\widetilde{B}_{1,1}-\widetilde{B}_{2,2})/2 \nonumber \\
\widetilde{U} & =(\widetilde{B}_{1,2}+\widetilde{B}_{2,1})/2 \nonumber \\
\widetilde{V} & =i(\widetilde{B}_{2,1}-\widetilde{B}_{1,2})/2 \nonumber .
\end{align}

For the randomly polarised sky, we expect $\widetilde{Q}=\widetilde{U}=\widetilde{V}=0$.

\subsection{Calibration and beam correction}
\label{sec:calib_and_beam_corr}

A standard calibration procedure is to observe a bright (dominating the visibilities), unresolved  and unpolarised 
source and solve for the complex gains of each tile via a least-square method. 
For an unpolarised calibrator source of intensity $I$, the sky brightness $B$ is given by 
\begin{equation}
\mathbf{B=}
\begin{bmatrix}
\begin{array}{cc}
1 & 0\\
0 & 1
\end{array}
\end{bmatrix}I,
\label{eq:b_upol_source}
\end{equation}
where we follow the ``Convention-1'' definition of Stokes $I$ \citep{Smi11-I} which was implemented in the  Common Astronomy Software Applications (CASA) \citep{casa}.
Assuming identical tile beams, eq.~\ref{eq:Vij} becomes

\begin{equation}
\fontsize{0.3cm}{1pt}\selectfont
\begin{split}
\mathbf{V}_{ij} = I \times \hspace{3.5cm} \\ 
\begin{bmatrix}
\begin{array}{cc}
  g_{i,x}(\lvert E_{x\theta}\rvert^{2}+\lvert E_{x\phi}\rvert^{2})g_{j,x}^{*} & g_{i,x}(E_{x\theta}E_{y\theta}^{*}+E_{x\phi}E_{y\phi}^{*})g_{j,y}^{*}\\ \\ \\ \\
  g_{i,y}(E_{y\theta}E_{x\theta}^{*}+E_{y\phi}E_{x\phi}^{*})g_{j,x}^{*} & g_{i,y}(\lvert E_{y\theta}\rvert^{2}+\lvert E_{y\phi}\rvert^{2})g_{j,y}^{*}
\end{array}
\end{bmatrix},
\label{eq:Vij-1-1}
\end{split}
\end{equation}

where $g_{x}$ and $g_{y}$ are the DIEs for the respective tile polarisations
and $\mathbf{E}(\theta,\phi)$ is the beam pattern at the ($\theta,\phi$)
direction of the calibrator source (we drop the $\theta,\phi$ notation
in the following equations). 
\red{Calibration solves for the $g_{x}$ values from the XX visibilities and likewise the $g_{y}$ values from the YY visibilities. 
The DDE are taken into account by either correcting the calibrator source model for the beam pattern $\mathbf{E}$ prior to solving for complex gains
(as implemented in \textsc{calibrate}) or by dividing the resulting complex gains by amplitudes of the electric field of X and Y dipoles 
( $(\lvert\tilde{E}_{x\theta}\rvert^{2}+\lvert\tilde{E}_{x\phi}\rvert^{2})^{1/2}$ and $(\lvert\tilde{E}_{y\theta}\rvert^{2}+\lvert\tilde{E}_{y\phi}\rvert^{2})^{1/2}$ respectively).}
\red{Both approches lead to DIE complex gains which can be represented as diagonal matrix:}


\begin{align}
\mathbf{\widetilde{G}} & =\begin{bmatrix}\begin{array}{cc}
\widetilde{g_{x}} & 0\\
0 & \widetilde{g_{y}}
\end{array}\end{bmatrix}.
\label{eq:complex-gain-estimate}
\end{align}

Note that we assume the complex gain matrix (eq.~\ref{eq:complex-gain-estimate}) to be diagonal, because off-diagonal terms are very small (negligible in comparison with mutual coupling of x and y dipoles) due to high isolation between the x and y analogue chains in MWA.

We introduce matrix $\mathbf{D}_{ij}$ as a specific instance of visibility data (dataset) obtained from some specific sky observation.
Subsequently, calibration of the observed visibility data $\mathbf{D}_{ij}$ corrects for the DIEs on the XX, YY, XY and YX polarisations:
\begin{equation}
\begin{split}
\mathbf{\widetilde{D}}_{ij} & =\mathbf{\mathbf{\widetilde{G}}}_{i}^{-1}\mathbf{D}_{ij}(\mathbf{\widetilde{G}}_{j}^{H})^{-1} = \\
 & \begin{bmatrix}\begin{array}{cc}
D_{ij}(0,0)/(\widetilde{g}_{i,x}\widetilde{g}_{j,x}^{*}) & D_{ij}(0,1)/(\widetilde{g}_{i,x}\widetilde{g}_{j,y}^{*})\\
D_{ij}(1,0)/(\widetilde{g}_{i,y}\widetilde{g}_{j,x}^{*}) & D_{ij}(1,1)/(\widetilde{g}_{i,y}\widetilde{g}_{j,y}^{*})
\end{array}\end{bmatrix}.
\end{split}
\label{eq:Dij-calibration}
\end{equation}
In the image space, this is our measurement of the apparent sky (eq.~\ref{eq:Bapp}), and from eq.~\ref{eq:B-est} we can calculate the sky brightness matrix as:
\begin{equation}
\mathbf{\widetilde{B}}=\mathbf{\widetilde{E}}_{{\rm obs}}{}^{-1}\widetilde{\mathbf{D}}\left(\mathbf{\widetilde{E}}_{{\rm obs}}^{H}\right)^{-1},\label{eq:B-est-meas}
\end{equation}
where $\tilde{\mathbf{E}}_{{\rm obs}}$ is our pointed beam model
in the direction of the observed target source. It is then trivial to use equation~\ref{eq:Stokes} to calculate Stokes parameters.

\subsection{Implementation of beam calibration pipeline}
\label{sec:beam_calib_pipeline}

We have three beam models to test:
\begin{itemize}
\item Analytic model of an array of \rednew{Hertzian} dipoles above a metallic ground plane \citet{SutOSu15}.
\item Average embedded element (AEE) model reported in \citet{SutOSu15}.
\item Full embedded element (FEE) model described in this paper.
\end{itemize}
\red{All three models have been implemented in MWA reduction software MWAtools, \textsc{rts} and \textsc{calibrate}.}
We process the same observations independently for each beam model. The steps are:
\begin{enumerate}
\item \red{Observe a calibrator source and use \textsc{calibrate} \citep{offringa-2016} to solve for the tile-based direction-independent complex gains, a diagonal matrix $\mathbf{\widetilde{G}}$ (eq.~\ref{eq:complex-gain-estimate}). The \textsc{calibrate} software \rednew{incorporates the beam model under test} into the calibration procedure.}
\item \red{Apply direction-independent calibration solutions to visibilities from target field.}
\item \red{Create sky images in all instrumental polarisations (XX,XY,YX and YY) with the WSCLEAN software \citep{OffMcK14}.}
\item Use the full-Stokes beam $\mathbf{\widetilde{E}}$, modelled for all ($\theta,\phi$) directions of the observation tile beam pointing (eq.~\ref{eq:B-est-meas}), to calculate the sky brightness matrix $\mathbf{\widetilde{B}}$ from the instrumental polarisations.
\item Calculate images in Stokes polarisation according to equation~\ref{eq:Stokes}.
\item Use the \textsc{Aegean} source finder \citep{aegean} to identify sources in Stokes I images, measure their flux densities in Stokes I, Q, U and V images and measure false Stokes Q, U and V relative to Stokes I.
\end{enumerate}

\section{FULL-STOKES DEMONSTRATION ON MWA DATA}
\label{sec_results}

In the following sections we will apply the above primary beam correction procedure to MWA data and compare performance of the three primary beam models. 
\red{In the first section we present comparison of the three models applied to the GLEAM data using our procedure as described in the previous section.
In Section~\ref{subsec:gleam_data_calibration} we summarise the original GLEAM calibration procedure and how the beam model was applied, then 
in Section~\ref{subsec:aplication_to_gleam_data} we present a three-night sample of GLEAM data re-calibrated with the FEE model and compare the resulting false Stokes Q between the AEE and FEE model.
Finally, in Section~\ref{subsec:gleam_expectations} we explain the false polarisation effect observed in the original GLEAM data calibrated with the AEE model.
}

\subsection{Comparison of models performance on MWA data}
\label{subsec:compare_models_mwa_data}

%
\red{To test the accuracy of the different beam models by measuring false leakages in all Stokes parameters (Q, U and V), we developed a pipeline implementing our full calibration procedure according to steps described in Section~\ref{sec:calib_and_beam_corr} and applied it to typical GLEAM observations at high frequencies (200-230\,MHz)}. 
These frequencies are most severely affected by inaccuracies of beam models, since some approximations of the physical model representation in FEKO become less accurate at shorter wavelengths (for example arms of the dipoles are represented as 15~mm\textbf{ } diameter wires instead of aluminium channels).
\red{At frequencies below 170 MHz the false Stokes polarisation is below 5\% for both AEE and FEE models.}
For calibration we used a 116\,s observation of Hydra~A starting at 13:24:48 UTC on 2014-03-06 (LST $\approx$8.14\,hours) at $(\phi,\theta) \approx (52.1\degree,22.0\degree)$.
The following steps were performed in order to probe the differences between the models:

\textbf{I. Solve for the tile-based direction-independent complex gains (Step 1 in Section~\ref{sec:beam_calib_pipeline})}

\red{As Hydra~A is partially resolved at MWA frequencies \citep{SutCol15,LanCla04}, we only use visibilities with baselines of length $30-300\lambda$ during calibration so that Hydra~A appears as an unresolved source. 
The \textsc{calibrate} software uses the beam model to correct the calibrator model. Hence, the resulting complex gains are already direction-independent complex gains.}

\textbf{II. Measure false Stokes on the calibration observation (Steps 2-6 in Section~\ref{sec:beam_calib_pipeline})}

\begin{figure*}
  \begin{center}
    \subfloat[Stokes I]{\includegraphics[height=2.82in]{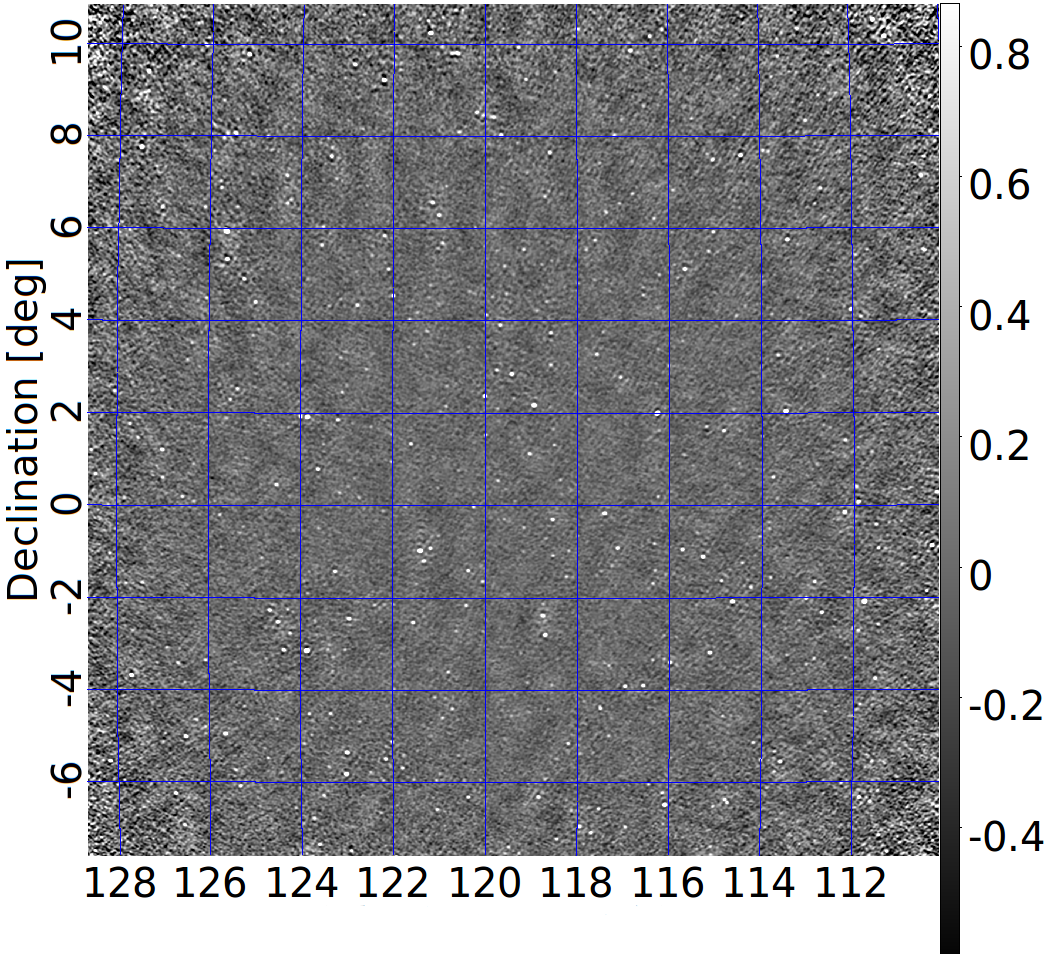}}
    \subfloat[Stokes Q]{\includegraphics[height=2.82in]{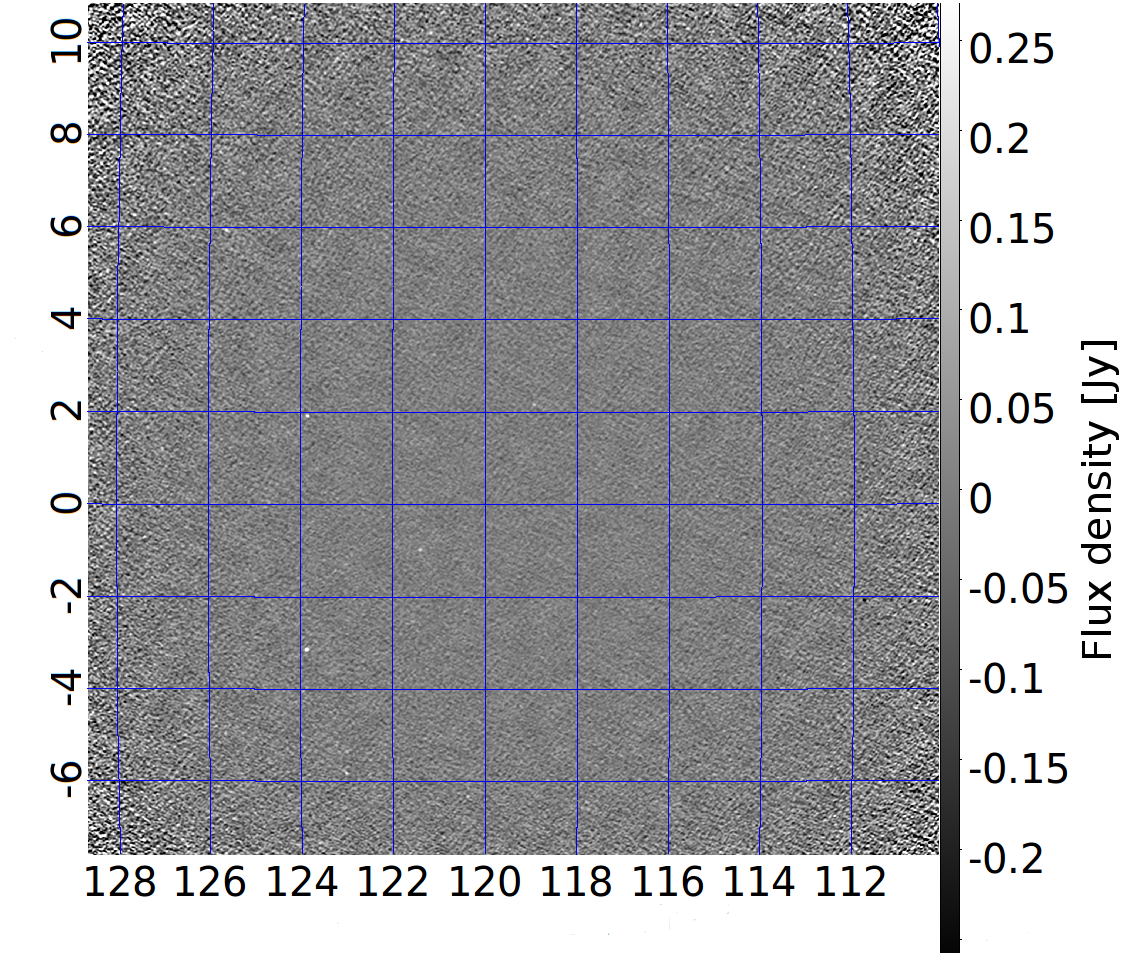}}\\
    \subfloat[Stokes U]{\includegraphics[height=2.82in]{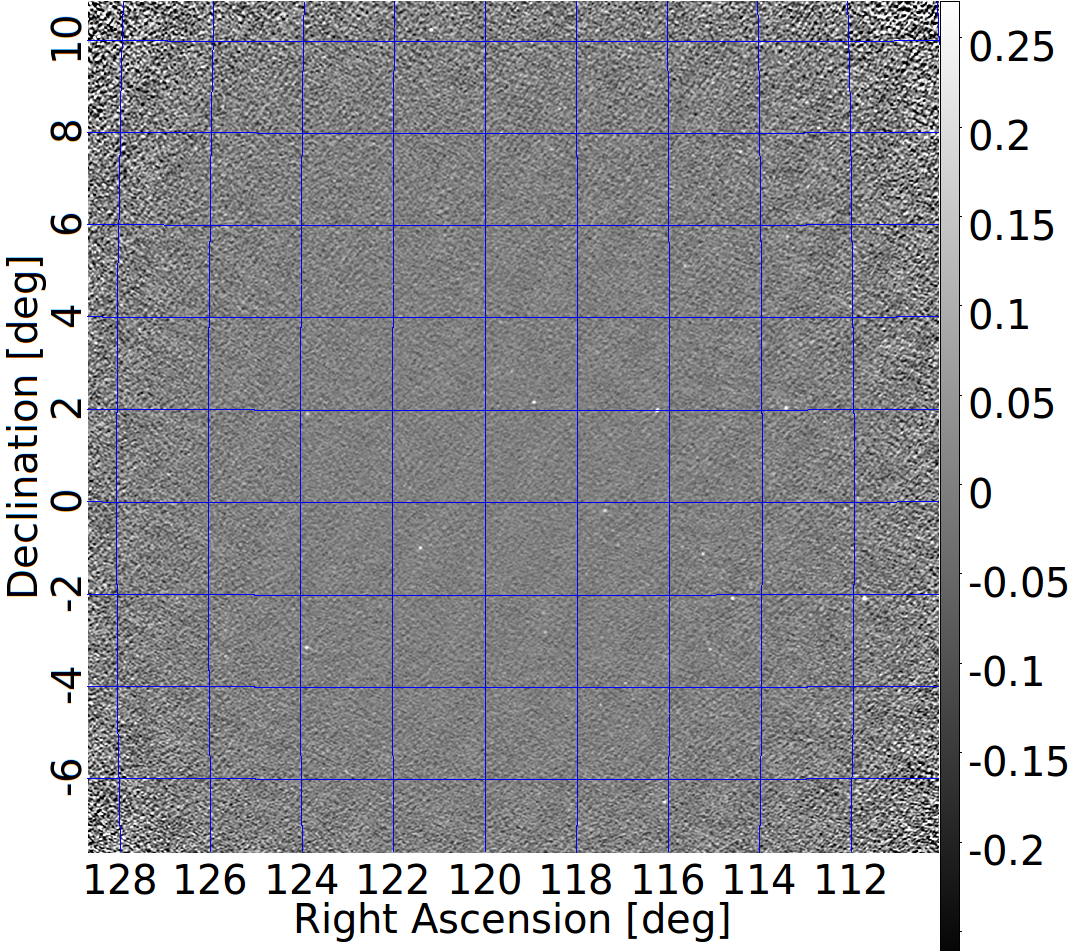}}
    \subfloat[Stokes V]{\includegraphics[height=2.82in]{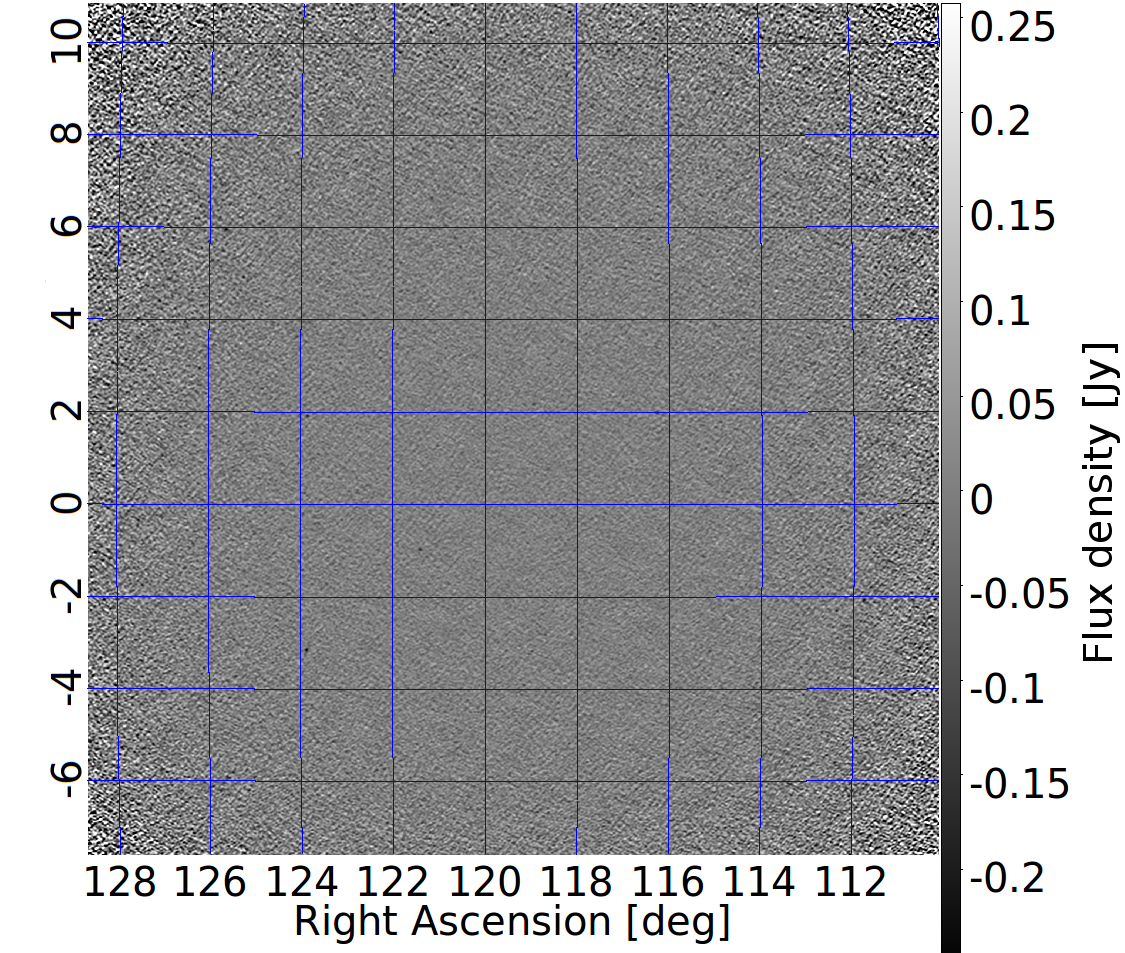}}
    \caption{Stokes I, Q, U and V images (a,b,c,d respectively) obtained from 2 minutes observation started at 13:14:48 UTC on 2014-03-06.
    \red{The images were beam corrected in 1.28\,MHz coarse channels and averaged in the 200-212\,MHz band. Only part of the band was used to avoid radio-frequency interference that affected the upper part of the band (most likely due to digitial TV) which caused subtle artefacts in the Q, U and V images. The images obtained with \textsc{wsclean} were beam corrected using the FEE model. The false Stokes leakages are within $\pm$5\% in image centres and get a bit higher closer to the edges. The $2^{nd}$ order polynomial surfaces fitted to false Stokes Q, U and V leakages are shown in Fig.~\ref{fig:2d_leakages} and leakages averaged in declination bins are show in Figures~\ref{fig:stokes_q_leakage_vs_dec_1078140304},~\ref{fig:stokes_u_leakage_vs_dec_1078140304}~and~\ref{fig:stokes_v_leakage_vs_dec_1078140304}.}}
    \label{fig:iquv_images}
  \end{center}
\end{figure*}

%
%
\begin{figure*}
  \begin{center}
    \subfloat[Q leakage]{\includegraphics[height=2.82in]{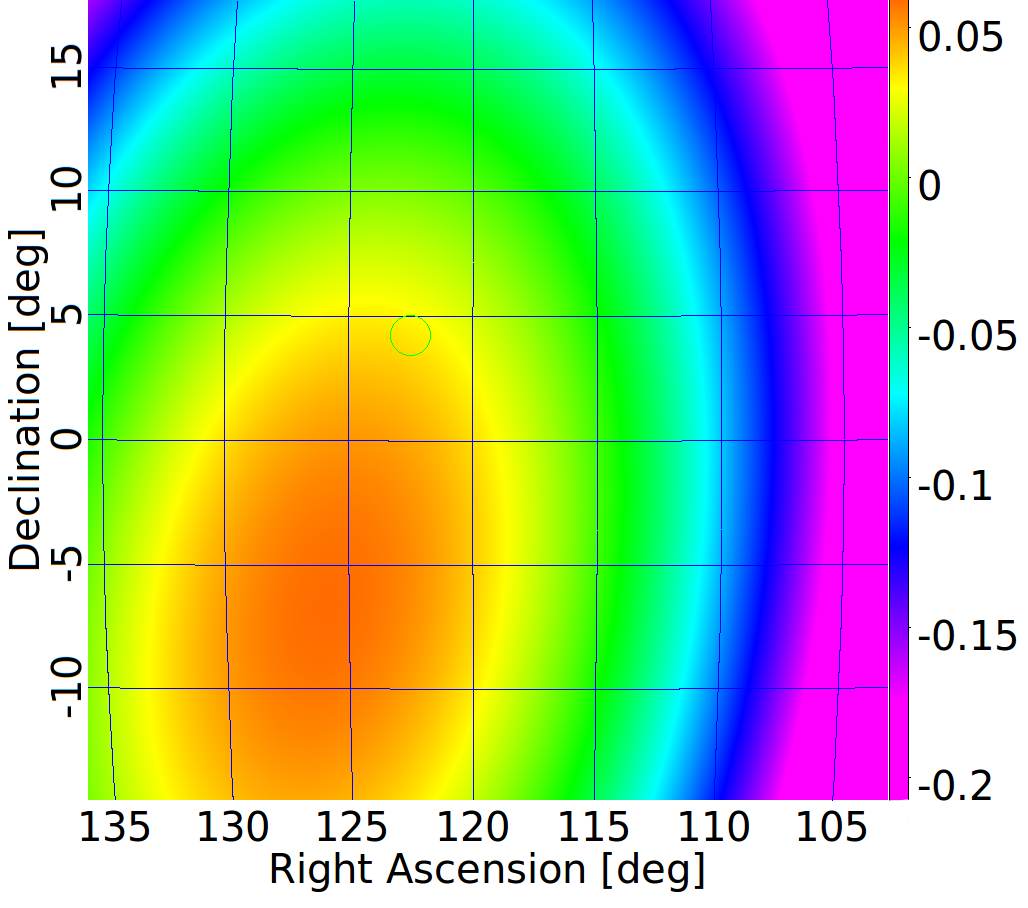}}
    \subfloat[U leakage]{\includegraphics[height=2.82in]{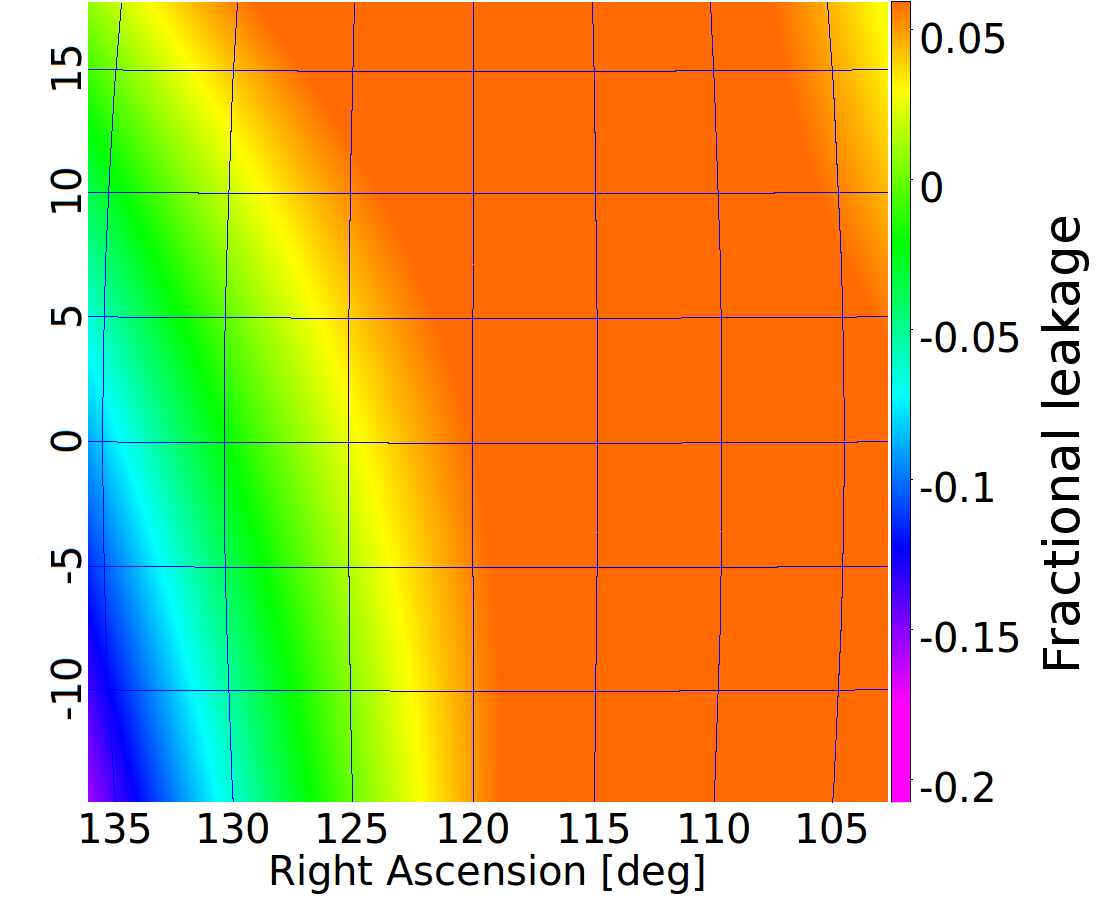}}\\
    \subfloat[V leakage]{\includegraphics[height=2.82in]{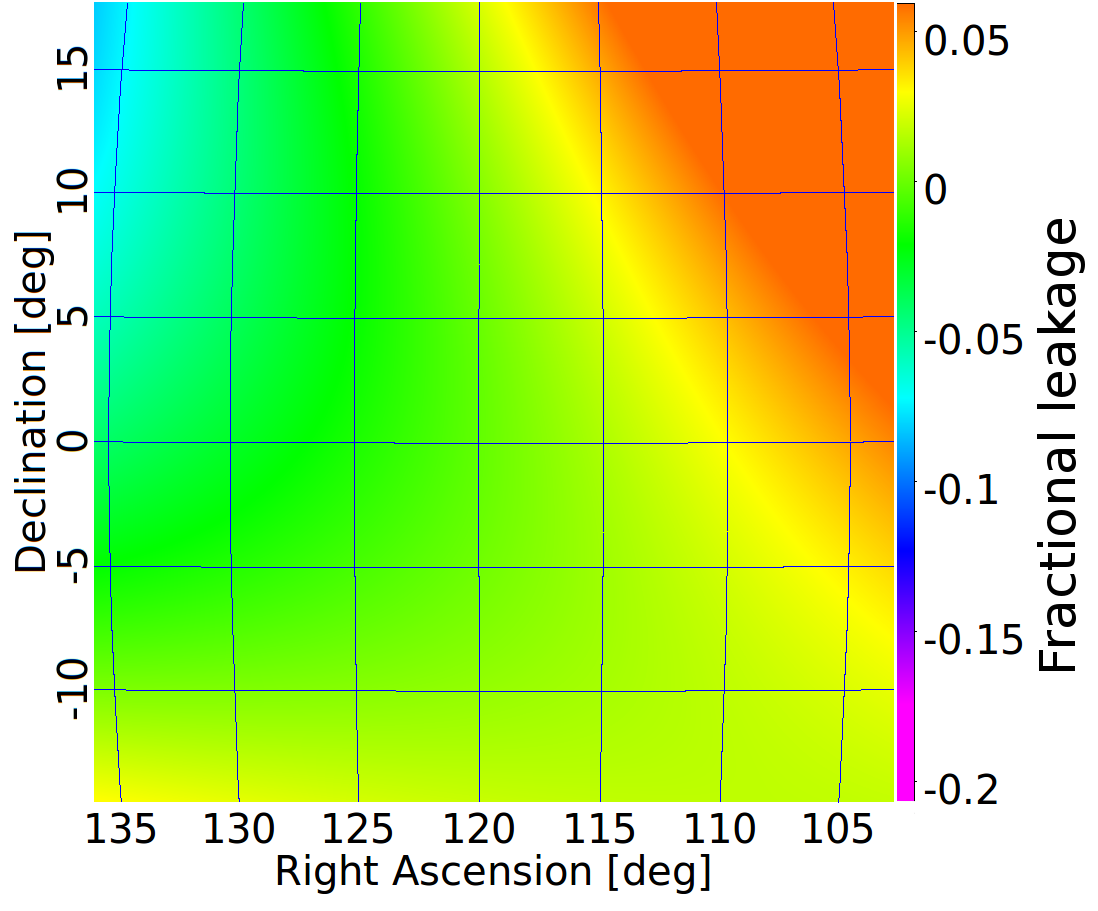}}
    \caption{False Stokes Q, U and V leakages (a,b,c respectively) surfaces obtained from fit of $2^{nd}$ order polynomial to leakages of the brightest sources from the Q, U and V images in Fig.~\ref{fig:iquv_images} \rednew{(the colour scale is the same for all three images)}.}
    \label{fig:2d_leakages}
  \end{center}
\end{figure*}

As described earlier, the beam model is applied at two stages: to correct calibration solutions for the beam response in the direction of the calibrator source and later to beam-correct an image of another field observation. Therefore, we verify the accuracy of the calibration correction at both stages. 
In the first check, we apply the calibration solutions to correct the visibilities of the calibration observation, estimate the sky brightness matrix and calculate the Stokes parameters. 
\red{The Stokes Q, U and V images of the calibrator fields are consistent with noise.}

\textbf{III. Transfer of calibration to another observation (Steps 2-6 in Section~\ref{sec:beam_calib_pipeline})}

\begin{figure*}
  \begin{center}
    \includegraphics[width=6in]{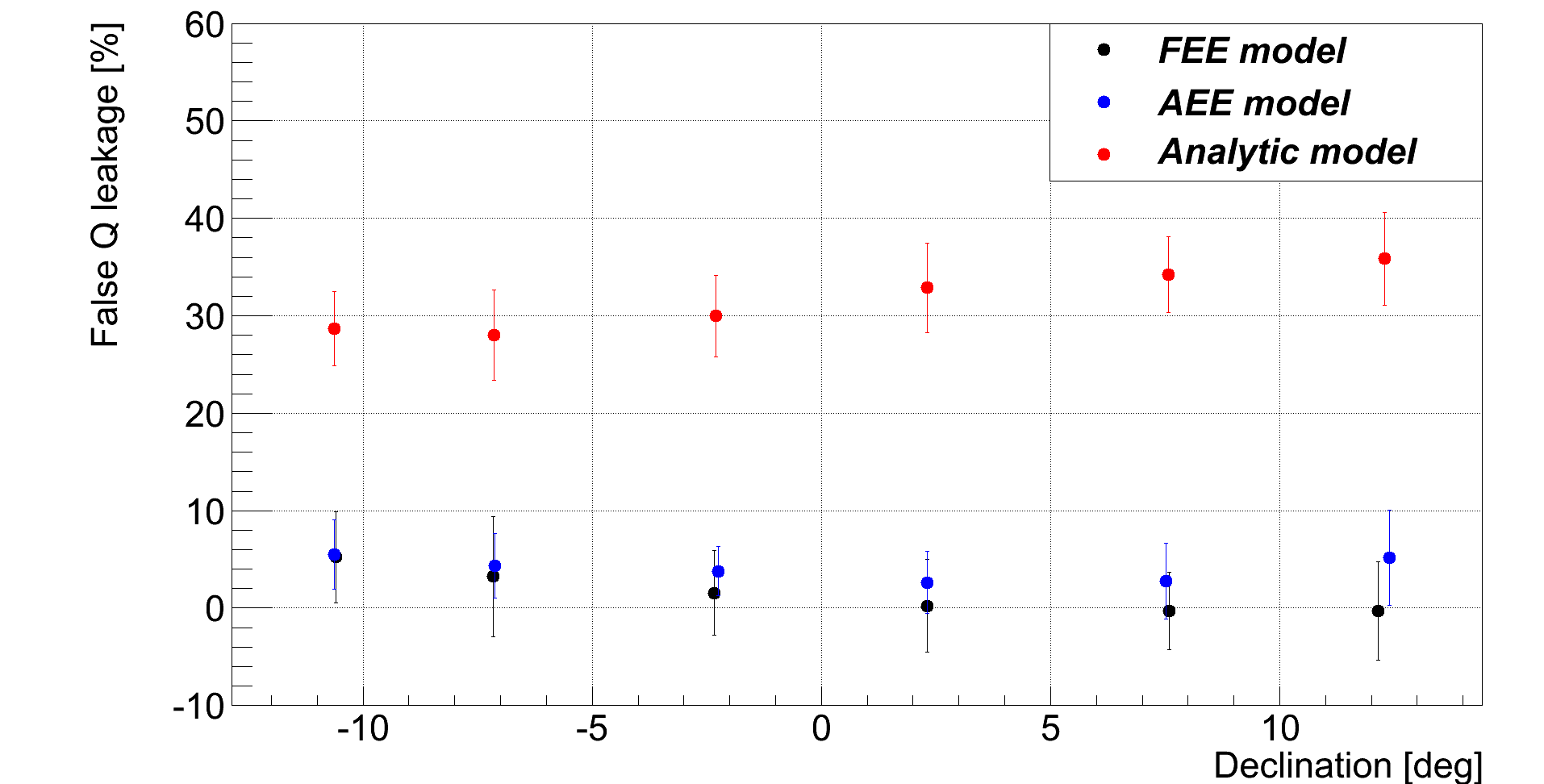}    
    \caption{\red{Comparison of Stokes Q leakage measured in images at 200--231\,MHz calibrated with the three different models. The leakage data from individual sources were averaged in 5$\degree$ bins (100-200 sources per bin). The analytic model performs the worst of all three models. The FEE models performs better at positive declinations and converges to AEE model at negative declinations (both are within measurements errors on this relatively small data sample).}}
    \label{fig:stokes_q_leakage_vs_dec_1078140304}
  \end{center}
\end{figure*}

\begin{figure*}
  \begin{center}
    \includegraphics[width=6in]{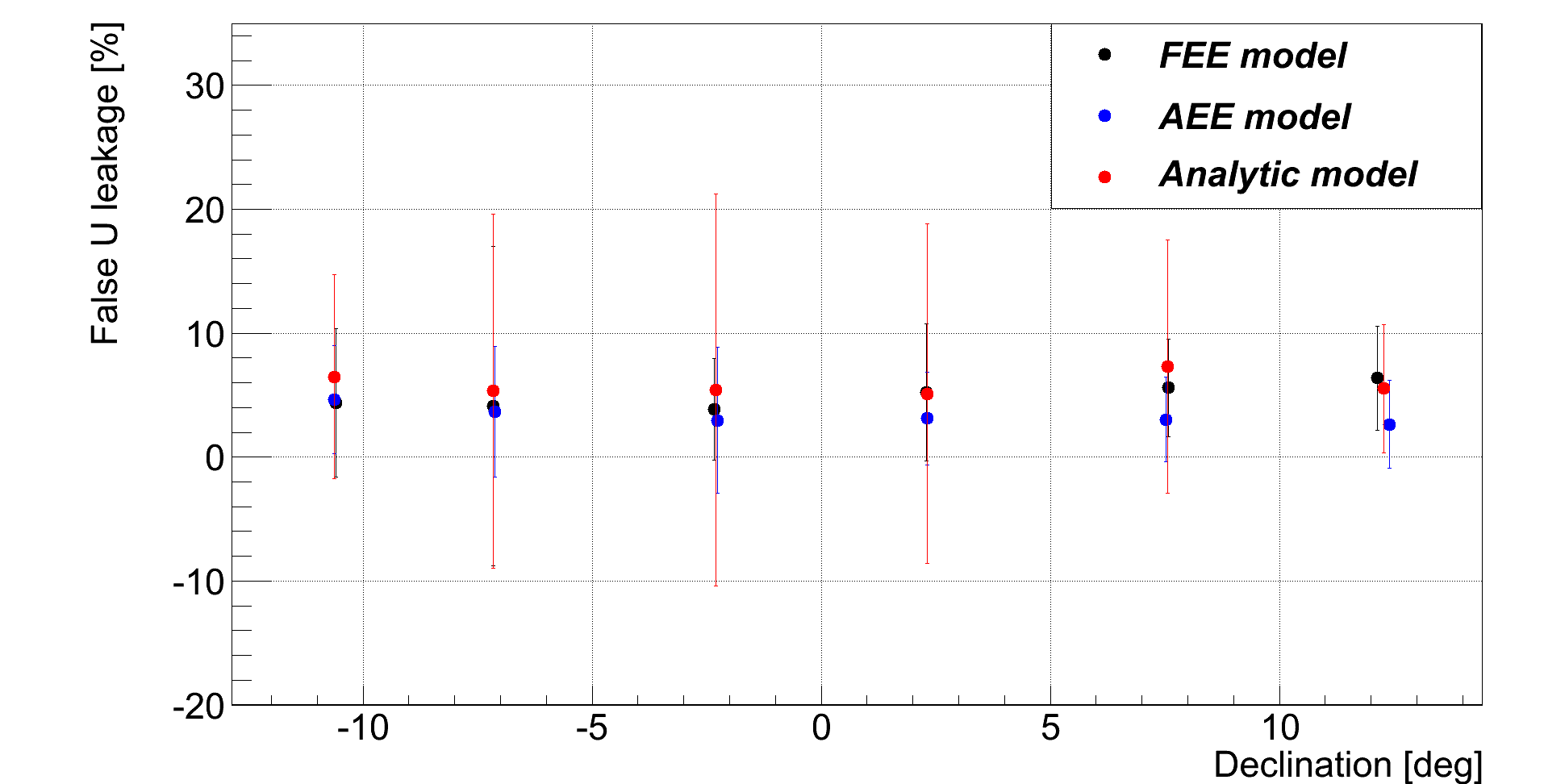}    
    \caption{\red{Comparison of Stokes U leakage measured in images at 200--231\,MHz calibrated with the three different models. The leakage data from individual sources were averaged in 5$\degree$ bins (100-200 sources per bin). All three models have similar values of the U leakage $\approx 5$\%. Note that the errors of false Stokes U for the analytical model are significantly larger than for the FEE and AEE models (Fig.~\ref{fig:stokes_u_leakage_vs_dec_1078140304}) because the hour angle dependence of false Stokes U was reduced significantly for the non-analytic models.}}
    \label{fig:stokes_u_leakage_vs_dec_1078140304}
  \end{center}
\end{figure*}

\begin{figure*}
  \begin{center}
    \includegraphics[width=6in]{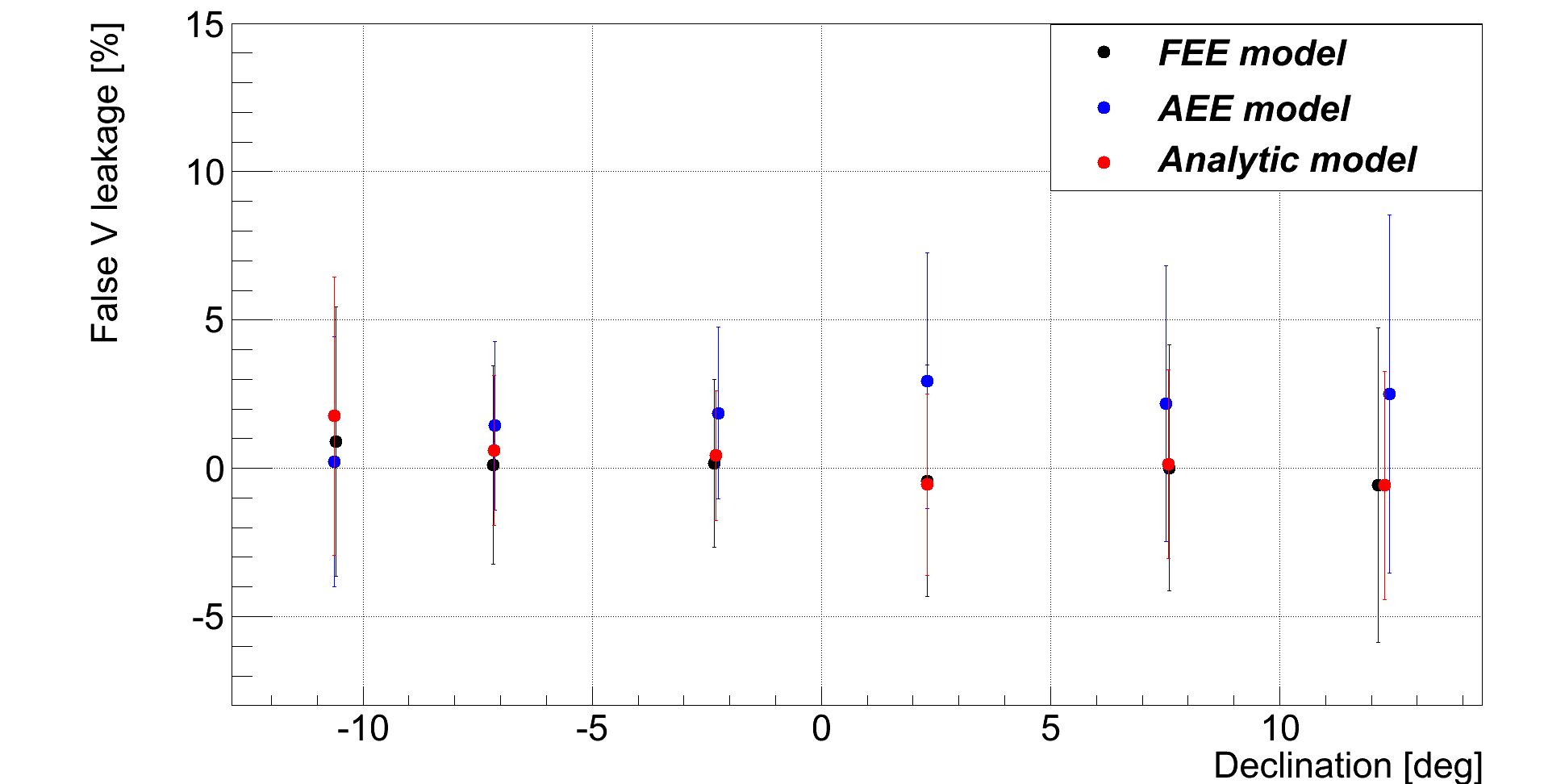}    
    \caption{\red{Comparison of Stokes V leakage measured in images at 200--231\,MHz calibrated with the three different models. The leakage data from individual sources were averaged in 5$\degree$ bins (100-200 sources per bin). All the models have values within the errors bars, but AEE model has V leakage $\approx$1-2\% consistent across the band, which is of similar magnitude to that reported by \citet{offringa-2016} below 200\,MHz.}}
    \label{fig:stokes_v_leakage_vs_dec_1078140304}
  \end{center}
\end{figure*}

In order to test beam correction in a typical observation scenario, we applied the direction-independent complex gains to a 
set of drift-scan observations at $(\phi,\theta) \approx (0\degree,28.3\degree)$ performed on 2014-03-06 between 11:44:47 and 13:16:40 UTC. \red{Stokes I, Q, U and V de-convolved images obtained with the \textsc{wsclean} and beam corrected with the FEE model are shown in Fig.~\ref{fig:iquv_images}. The corresponding $2^{nd}$ order polynomial surfaces fitted to the false Q, U and V leakages of the brightest sources are shown in Fig.~\ref{fig:2d_leakages}.}
\red{The false Stokes Q, U and V averaged in 5$\degree$ declination bins are shown in Figures~\ref{fig:stokes_q_leakage_vs_dec_1078140304},~\ref{fig:stokes_u_leakage_vs_dec_1078140304}~and~\ref{fig:stokes_v_leakage_vs_dec_1078140304} respectively.}

\red{The new FEE model has false Stokes Q below 5\% falling down from $\approx$5\% to 0\% with increasing declination (Fig.~\ref{fig:stokes_q_leakage_vs_dec_1078140304},~\ref{fig:stokes_u_leakage_vs_dec_1078140304}~and~\ref{fig:stokes_v_leakage_vs_dec_1078140304}). The data calibrated with the AEE model has slightly higher false Stokes Q, but also within 5\%. Both FEE and AEE models are better than the analytic model which has a false Stokes Q $\approx$30\% (Fig.~\ref{fig:stokes_q_leakage_vs_dec_1078140304}).
The false Stokes U is $\approx$5\% for all three models and the false Stokes V in the data corrected with FEE model is consistent with 0\%, whilst the V leakage in the data calibrated with the AEE model is $\approx$1-2\% (of similar magnitude to that reported by \citet{offringa-2016} below 200\,MHz). }

\red{However, the errors on this relatively small data sample are quite high and the results from all three models agree within the errors (except the false Stokes Q of the analytic model). 
These errors result from hour angle (HA) dependence of false Stokes leakage across the image (Fig.~\ref{fig:2d_leakages}), which is averaged in declination bins and the errors calculated as standard deviation correspond to variation in HA. 
Note that the errors of false Stokes U for the analytical model are significantly larger than for the FEE and AEE models (Fig.~\ref{fig:stokes_u_leakage_vs_dec_1078140304}) because the HA dependence of false Stokes U was reduced significantly for the non-analytic models.
In the next section we will show the effects of the new model on a larger (three nights) GLEAM data sample and how it improves the false Stokes Q originally observed in the calibrated GLEAM data.
}

%

\subsection{Original GLEAM calibration procedure}
\label{subsec:gleam_data_calibration}

\red{One of the goals of the GLEAM survey was to catalogue the flux density of all radio sources below $+30\degree$ declination in the 72-231\,MHz frequency band, but polarisation measurements were not initially a priority.
However, because beam-corrected instrumental XX and YY images have been calibrated independently to the Molonglo Reference Catalogue (MRC) catalogue \citep{1981MNRAS.194..693L} it was identified that the ratio of YY and XX fluxes deviate from unity (equivalent to non-zero false Stokes Q) away from the image centres.
The ratio has noticeable structure as a function of declination (sources were grouped in bins in declination and frequency as shown in Fig.~4 in ~\citet{gleam_nhw}).
The observed false Stokes Q were attributed by the authors to deficiencies in the AEE primary beam model, as the structure of YY/XX ratio versus declination remains the same between different nights (sometimes separated by three months), which excludes the possibility of the effect being due to variations in the ionosphere.}

\red{
In the next section we will show how application of the new FEE beam model reduces the false Stokes Q observed in a sample of GLEAM data taken over three nights.
Inaccuracies in the AEE model (which was used for GLEAM processing) resulted in flux-scale variations as a function of beam pointing and declination. 
To overcome this, a robust flux calibration procedure was developed for the GLEAM pipeline, as outlined in ~\citet{gleam_nhw}. }

\red{The GLEAM pipeline only provided calibrated XX and YY images and so beam model effectiveness can only be tested against Stokes Q. 
We used the new FEE beam model in the original GLEAM calibration pipeline in order to verify if it corrected the originally reported false Stokes Q.
}
\begin{figure*}
  \begin{center}
    \includegraphics[width=6in]{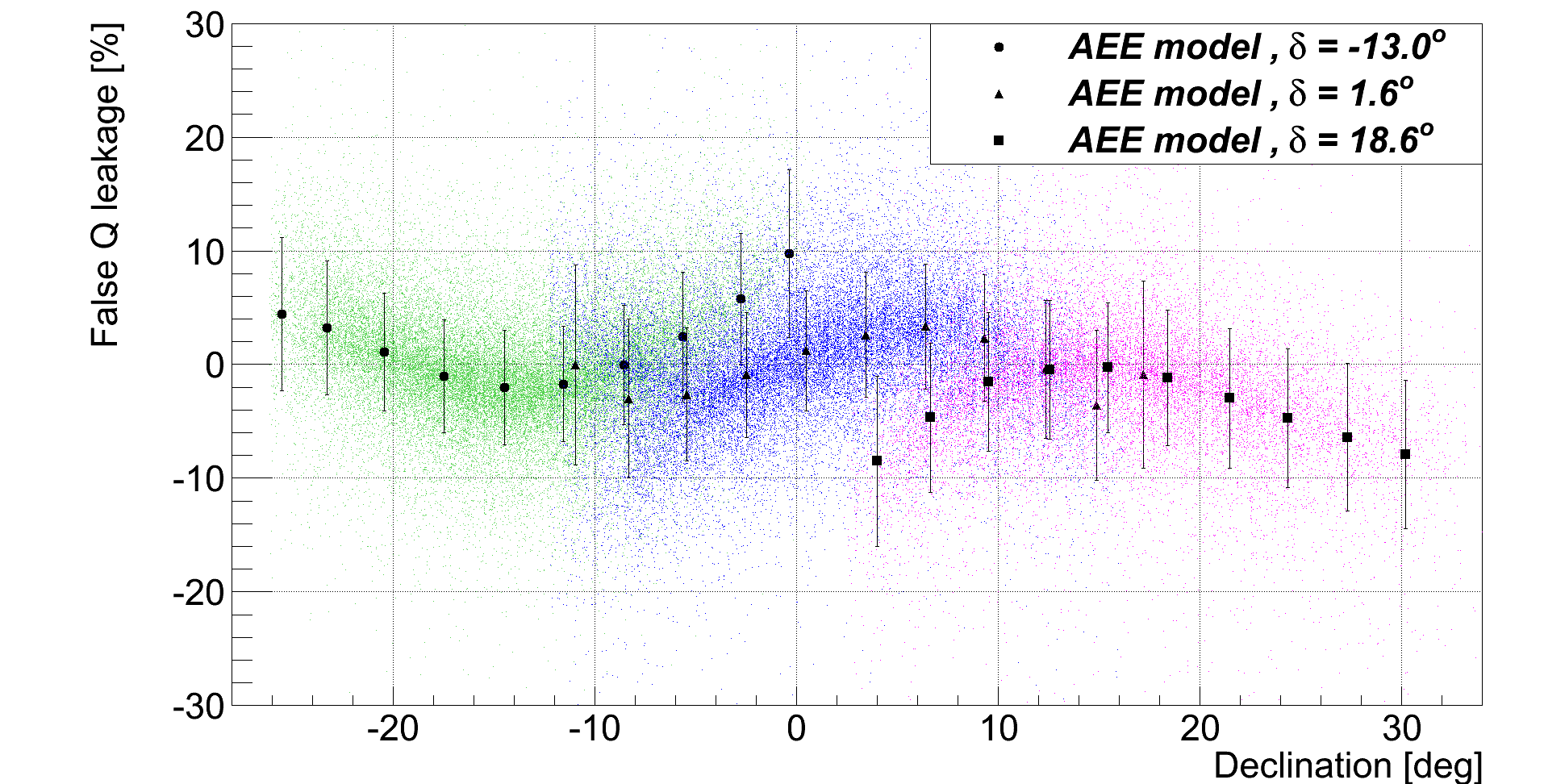}
    \includegraphics[width=6in]{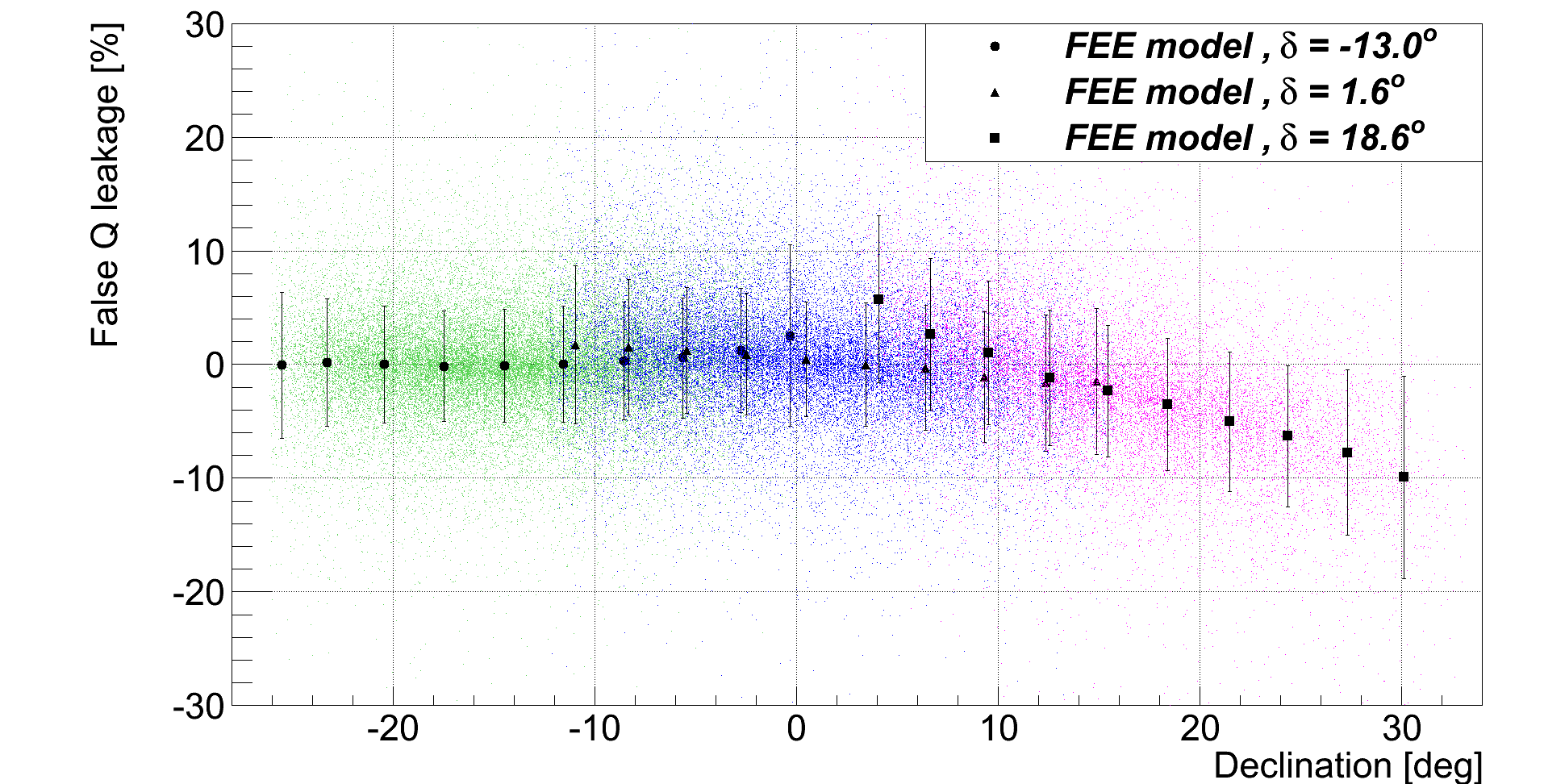}
    \caption{\red{False Stokes Q leakages calculated from GLEAM images corrected with the originally used AEE model (upper image) and the new FEE model (lower image). The small scattered data points represent false Stokes Q calculated for all the individual sources (around 85000 in total in both cases) identified in the images of the three fields at declinations $\delta= -13\degree, 1.6\degree$ and $+18.6\degree$ (with approximately 40100, 29000 and 16000 sources respectively). The large data points with error bars were calculate as the mean and standard deviation in 3$\degree$ bins.}}
    \label{gleam_leakage_aee_vs_fee}
  \end{center}
\end{figure*}

\subsection{Application to the GLEAM data}
\label{subsec:aplication_to_gleam_data}

\red{We applied the new FEE beam model to the GLEAM data from three pointings at the local meridian and declinations $\delta= -13\degree, +1.6\degree$ and $+18.6\degree$ collected during nights starting on the 5th, 7th and 11th of Nov 2013 respectively. 
The GLEAM data were processed as described in ~\citet{gleam_nhw}, but instead of using AEE model the new FEE beam model was applied.
We analysed images in four 7.68\,MHz bands (200--208\,MHz, 208--216\,MHz, 216--223\,MHz and 223--231\,MHz).
The false Stokes Q leakages were measured for many sources (nearly 85000 in total, with $\approx$40100 at $\delta=-13\degree$, $\approx$29000 at $\delta=1.6\degree$ and $\approx$16000 at $\delta=18.6\degree$) using \textsc{Aegean} (step 6 in Sec.~\ref{sec:beam_calib_pipeline}).}

\red{The comparison of false Stokes Q leakages by applying the AEE (as originally used in the GLEAM survey) and FEE models is shown in Fig.~\ref{gleam_leakage_aee_vs_fee}.
The false Stokes Q leakages resulting from calibration with the AEE model (upper plot in Fig.~\ref{gleam_leakage_aee_vs_fee}) have significant declination-dependent structure (reaching values around and above 10\% at the edges of the images) as a function of declination for each of the analysed fields. } 
In contrast, sources in the images obtained from calibration with the new FEE model have smaller false Stokes Q, which is consistent with zero for the two pointings ($\delta=-13\degree$ and $\delta=+1.6\degree$) and non-zero (within $\pm 10\%$) for the $\delta=18.6\degree$, which is at the lowest elevation ($\approx 45\degree$).
\rednew{Moreover, only the lowest elevation has a noticeable structure as a function of declination.}

%
\begin{figure*}
\begin{center}
\subfloat[200 - 208 MHz]{\includegraphics[width = 3.4in]{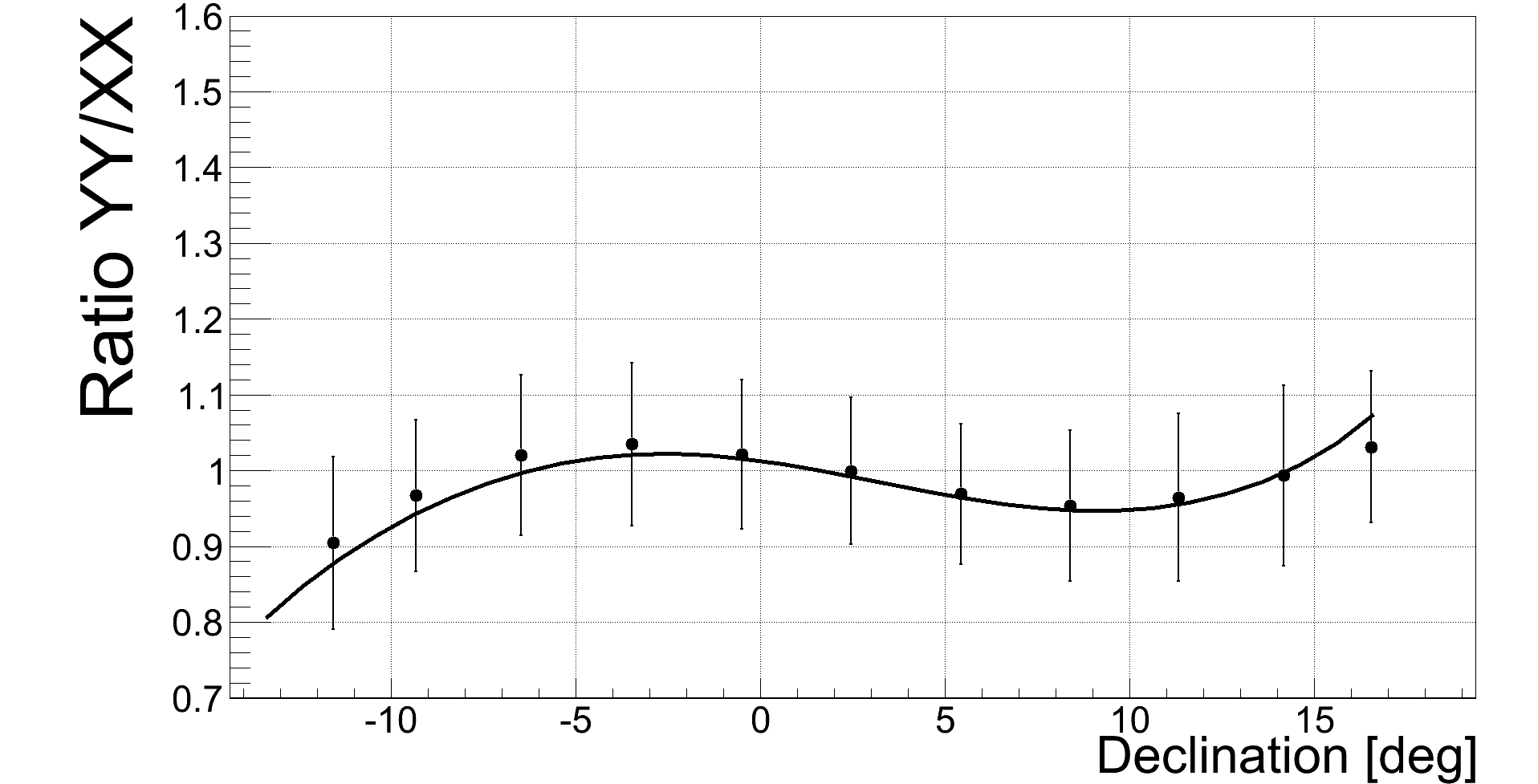}}
\subfloat[208 - 216 MHz]{\includegraphics[width = 3.4in]{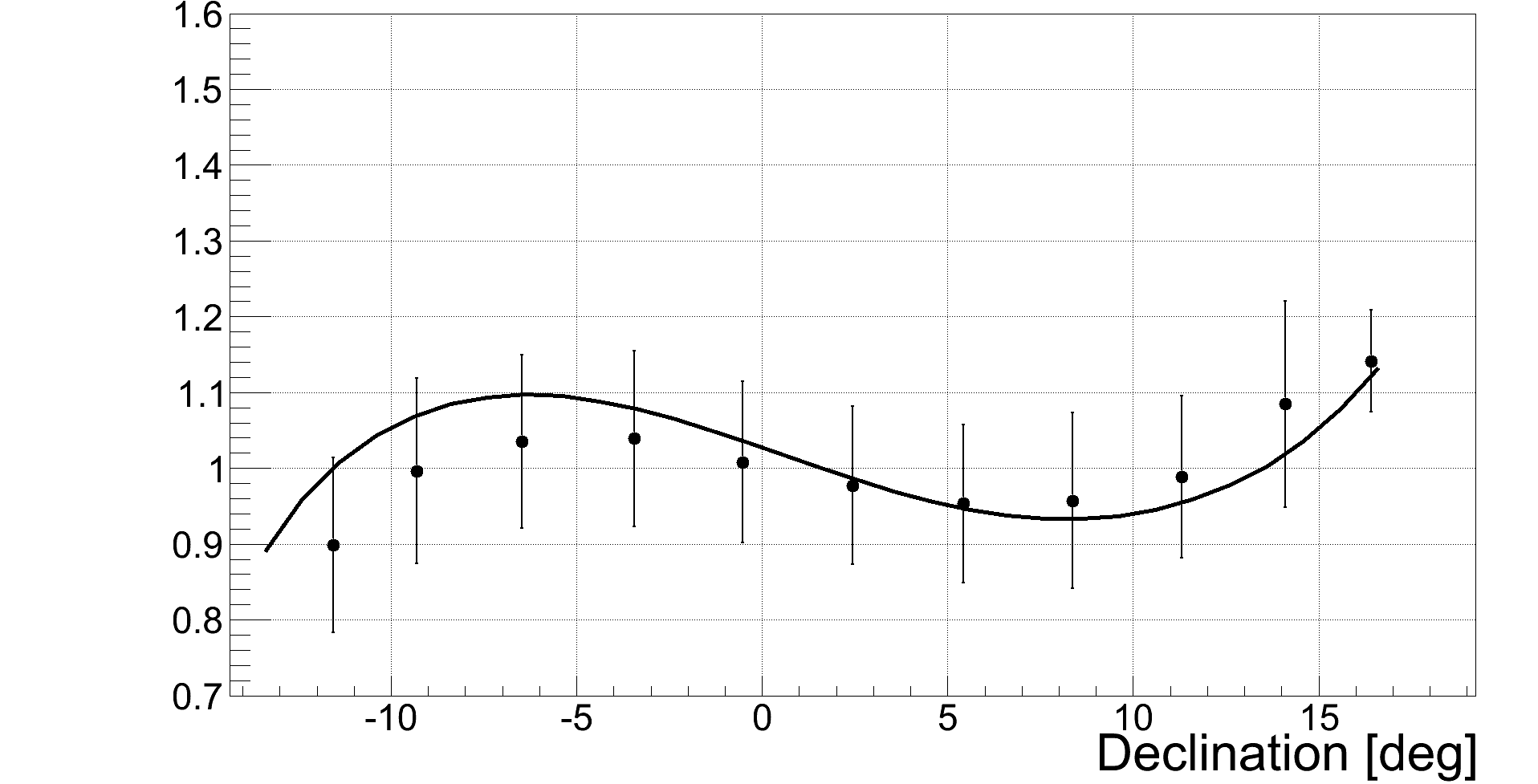}}\\
\subfloat[216 - 223 MHz]{\includegraphics[width = 3.4in]{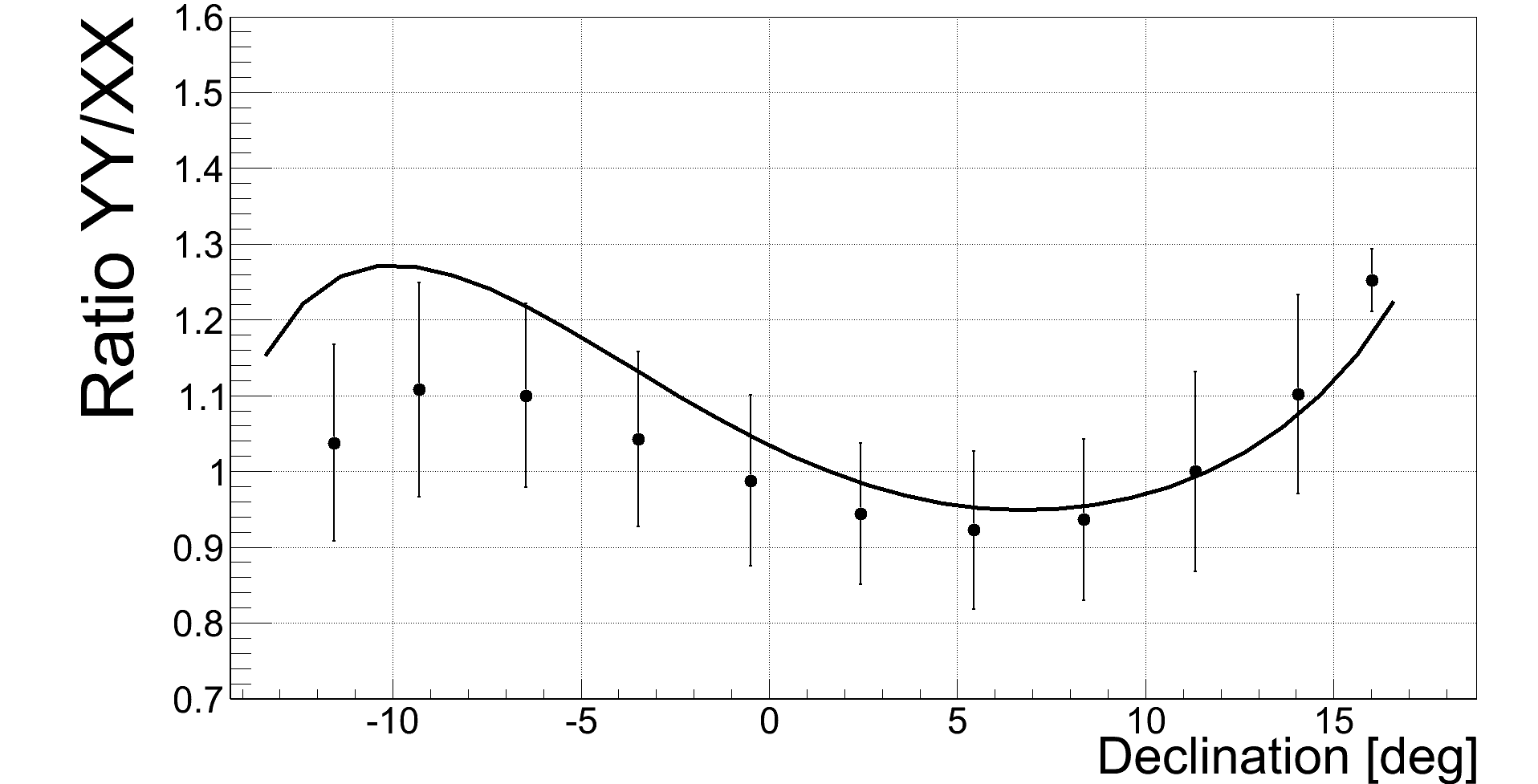}}
\subfloat[223 - 231 MHz]{\includegraphics[width = 3.4in]{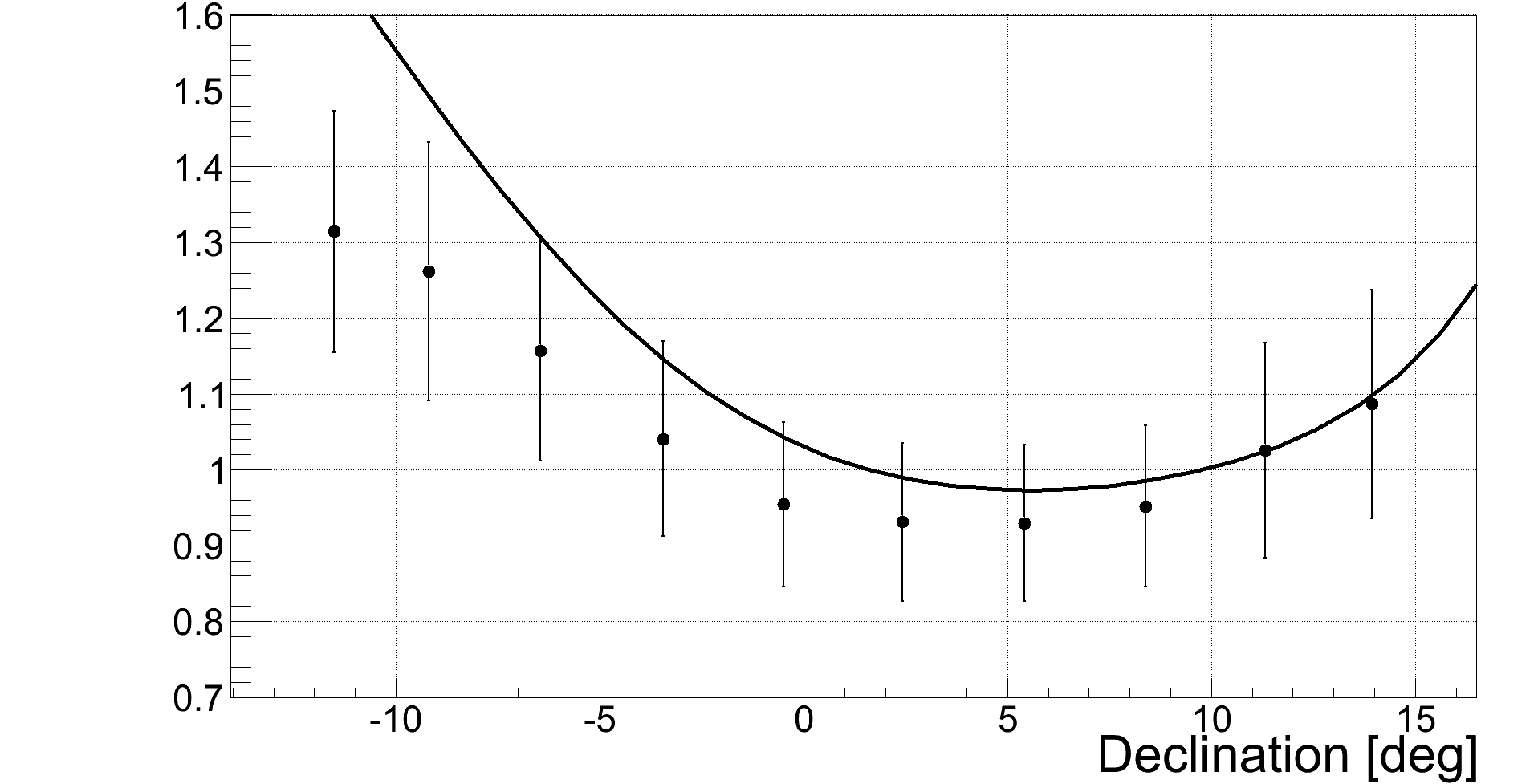}}
\caption{\red{The data points are the ratio $\widetilde{B}_{2,2} / \widetilde{B}_{1,1}$ based on the GLEAM data from bottom row (frequency ranges 200-208\,MHz, 208-216\,MHz, 216-223\,MHz and 223-231\,MHz) of Fig.4 in \citep{gleam_nhw}. The data were collected with the beam pointing at local meridian at declination $\approx 2 \degree$. The GLEAM data were binned in $3 \degree$ bins in declination (with about 1000-1200 sources averaged in the central bins down to 100-200 in the bins near the image edges). The solid lines represent the same ratio predicted when the AEE beam model is used to correct unpolarised sky brightness propagated through the FEE beam model (assumed to represent the ``true'' MWA tile beam). The simulated curves were normalised by values in the image centre in order to replicate the normalisation of flux to The Molonglo Reference Catalogue (MRC) catalogue \citep{1981MNRAS.194..693L} performed on the GLEAM data.}}
\label{fig:ratio_Sxx_div_Syy}
\end{center}
\end{figure*}

\subsection{False Stokes Q expected in GLEAM due to inaccuracy of the beam model}
\label{subsec:gleam_expectations}

\red{The GLEAM survey identified that the ratio of YY/XX flux densities deviates from unity away from the pointing centres and has a declination-dependent structure as shown in Fig.~4 in \citep{gleam_nhw} and Fig.~\ref{fig:ratio_Sxx_div_Syy} in the current paper (only the data $>200$\,MHz). 
The effect was reduced, but still noticeable, by the mosaicking procedure which up-weighted (by square of the beam pattern $E^2$) the measurements taken closer to the beam centre (local meridian). 
Using the YY/XX ratios observed in GLEAM we calculated the equivalent in terms of false Stokes Q, which is in a range of approximately $5-20\%$.
} 

\red{Because of the way the GLEAM data were processed, i.e. images in XX and YY instrumental polarisations were beam corrected and their flux densities normalised independently to the MRC catalogue, the observed structure was attributed to inaccuracies of the AEE beam model (ionosphere was excluded by intra-night stability of the declination structure). 
Therefore, with the new FEE model available, we were able to verify if the observed declination structure could be reproduced by assuming a priori the FEE beam model to be a ``true'' (or a better representation) of tile beam and applying the AEE model to calibrate the data. In this test we only tested beam related effects and did not take electronic DIE gain into account.
We assumed that the unpolarised sky brightness $\mathbf{B_{unpol}}$ is given by equation~\ref{eq:b_upol_source} and hence we calculated the ``calibrated'' sky brightness according to the following equation:}
\begin{equation}
\widetilde{\mathbf{B}} = \mathbf{E_{AEE}}^{-1} \Big[ \mathbf{E_{FEE} B_{unpol} E_{FEE}^H} \Big] \left( \mathbf{E_{AEE}}^H \right)^{-1}.
\label{eq:daniel_test}
\end{equation}

\red{The comparison between the ratio YY/XX observed in the original GLEAM data (frequency bands 200--231\,MHz in Fig. 4 in \citet{gleam_nhw}) and our prediction for the same quantity (calculated as $\widetilde{B}_{2,2} / \widetilde{B}_{1,1}$) is shown in Fig.~\ref{fig:ratio_Sxx_div_Syy}.}
\red{The GLEAM authors attributed the declination-dependent structure in the ratio YY/XX as a result of an inaccurate beam model. Our predictions, shown in Fig~\ref{fig:ratio_Sxx_div_Syy}, support this statement.}
\section{CONCLUSIONS AND DISCUSSION}
\label{sec:dicussion}

An accurate primary beam model of the telescope is required to pursue many science goals of the MWA and the future SKA-low telescope.
\red{We have presented a new Full Embedded Element (FEE), which is the most rigorous realisation of the beam model for the MWA, superseding the previous Average Embedded Element (AEE) and analytical beam models \citep{SutOSu15}.}
The new model was generated in the FEKO software using an improved physical representation of MWA tile.
In the simulation every dipole in an MWA tile was simulated in transmit mode with other dipoles not transmitting (voltages set to zero).
\red{The FEKO simulation results were exported using spherical harmonic representation, which enables calculation of the beam pattern and Jones matrices in arbitrary pointing directions without re-running the simulations (instead of discrete directions as in the previous AEE model).}

\red{We used polarisation measurements to compare the three beam models. We have developed a beam calibration pipeline to calibrate the direction dependent and independent effects in MWA observations and measure false Stokes Q, U and V as a metric for accuracy of the models.
We applied this procedure to a set of $12\times2$-minute GLEAM observations. The analytical model shows very high ($\sim 30$\%) Q leakage, considerably higher than the measured leakage in both the AEE and FEE models.
The new FEE model has false Stokes Q leakage $0 - 5\%$. The AEE model gives higher absolute value of leakage at higher declinations and similar at lower declinations. 
The false Stokes U is similar for all three models (within $\pm 5$\%). The V leakage resulting from the FEE model is consistent with zero whilst the V leakage resulting from AEE models is $\approx$1-2\% (of similar magnitude to that reported by \citet{offringa-2016} below 200\,MHz).
However, they are both within measurement errors on the relatively small data sample we used for this test. The pipeline enables further tests of false Stokes leakages in all pointing directions in order to identify further avenues to improve the model.
}

\red{We have also applied the FEE model to the original GLEAM calibration pipeline (Stokes Q only) to correct three nights of GLEAM data and calculated false Stokes Q for nearly 85000 sources in these images. 
The FEE model reduces the declination-dependent structure of false Stokes Q in comparison to the AEE model and the false Stokes Q is consistent with zero in two out of three pointing directions.}

\red{We used the new FEE model to understand the Q leakage (and its structure as a function of declination) observed in the original GLEAM data. Using the new model as a ``hypothetical true'' MWA beam model and calibrating it with the previous (AEE) beam model we were able to reproduce the declination-dependent structure in false Stokes Q observed in GLEAM survey (\citet{GLEAM_Wayth}, \citet{gleam_nhw}). } 

\rednew{Although, the current model is the best representation of the tile beams yet,} there are still possibilities for further improvements in the physical model of the MWA tile by incorporating finer details into the simulation. However, the better physical representation of the physical tile the longer the simulation takes.
Therefore, the accuracy of the physical model is a trade-off between the required precision and practical constraints (for instance simulation time). 

Another possibility is that the observed false Stokes leakages result from deviation of the actual MWA telescope from the ``ideal'' instrument represented by the model.
Although most of the component variations should average to zero and not contribute significantly to differences between XX and YY polarisations causing false Stokes, some effects may not cancel out. 
We performed simulations and generated 128 beams taking into account information about faulty dipoles (which were disabled at beamformers), calculated the mean beam and performed a test similar to the one described in Section~\ref{subsec:gleam_expectations} (treating the mean beam as the ``true'' MWA beam and beam-corrected with the ideal model beam).
\red{Results of these tests showed that faulty dipoles (usually in one polarisation), especially when fault distribution is non-uniform within tiles, can lead to false Stokes leakage of the order similar to that observed (as just one faulty or disabled dipole represents 1/16 ($\approx$6\%) of the MWA tile ). 
In future, this effect can be tested with observed data. Any residual leakage as a result of inaccuracies in the physical model can be reduced by measuring, mapping and subtracting the residual leakage on a per snapshot basis as described by \citet{2017arXiv170805799L}.
}

Finally, it should be noted that electro-magnetic simulations have certain limitations and the resulting models are loaded with error (due to imperfection of the physical representation, variations of the components, numerical limitations etc.).
\red{The acceptable error of the simulation is considered to be of the order of a few percent which also corresponds to typical ($\sim0.2-0.4$\,dB) spread in RF component parameters. The currently used simulation resolution is already optimal as we have verified that increasing the mesh resolution by factor of two results in $\lesssim$1\% difference in the beam response, but at a significantly higher computing cost (factor of ~12).
The false Stokes leakages at the level of a few percent are close to current limitations of the electro-magnetic simulations, highlighting the fact that astronomical polarisation measurements offer a very sensitive and rigorous ways of testing and validating electro-magnetic simulations.}



\begin{acknowledgements}
The authors would like to thank the anonymous referee for their valuable input and suggestions, which have significantly improved the paper.
The International Centre for Radio Astronomy Research (ICRAR) is a Joint Venture between Curtin University and the University of Western Australia, funded by the State Government of Western Australia and the Joint Venture partners.
This scientific work makes use of the Murchison Radio-astronomy Observatory, operated by CSIRO. We acknowledge the Wajarri Yamatji people as the traditional owners of the Murchison Radio-astronomy Observatory site.
Support for the operation of the MWA is provided by the Australian Government (NCRIS), under a contract to Curtin University administered by Astronomy Australia Limited. 
We acknowledge the Pawsey Supercomputing Centre which is supported by the Western Australian and Australian Governments.
This research was undertaken with the assistance of resources from the National Computational Infrastructure (NCI), which is supported by the Australian Government.
The Centre of Excellence for All-sky Astrophysics (CAASTRO) is an Australian Research Council Centre of Excellence, funded by grant CE110001020.
\end{acknowledgements}

\nocite*{}
\bibliographystyle{pasa-mnras}
\bibliography{references}

\end{document}